\documentclass{aa}
\usepackage[varg]{txfonts}
\usepackage{lineno}
\usepackage{amsmath}
\usepackage{color}
\usepackage{multirow}
\usepackage{soul}
\usepackage[normalem]{ulem}

\graphicspath{{Figures/}}

\titlerunning{A discrete-time autoregressive model}
\authorrunning{Elorrieta et al.}

\usepackage{comment}

\usepackage{graphicx}

\begin{document}

\title{Discrete-time autoregressive model for unequally spaced time-series observations}

\author{Felipe Elorrieta\inst{1,3}\and
Susana Eyheramendy\inst{2,3,4,5}\and
Wilfredo Palma\inst{2,3}}   

\institute{Departmento de Matem\'aticas, Facultad de
  Ciencia, Universidad de Santiago de Chile, Av. Libertador Bernardo 
  O'Higgins 3663. Estacion Central, Santiago, Chile
\and Departmento de Estad\'istica, Facultad de
  Matem\'aticas, Pontificia Universidad
  Cat\'olica de Chile, Av.\ Vicu\~na Mackenna
    4860, 7
820436 Macul, Santiago, Chile
\and Millennium Institute of Astrophysics, Santiago, Chile
\and Max-Planck-Institut f\"ur Astronomie, Heidelberg, Germany
\and Faculty of Engineering and Sciences, Universidad Adolfo Iba\~nez, Diagonal Las Torres 2700, Pe\~nalol\'en, Santiago, Chile
}

\abstract{
Most  time-series models assume that the data come from observations that are equally spaced in time. However, this assumption does not hold in many diverse scientific fields, such as astronomy, finance, and climatology, among others. There are some techniques that fit unequally spaced time series,  such as the continuous-time autoregressive moving average (CARMA) processes. These models are defined as the solution of a stochastic differential equation. It is not uncommon in astronomical time series, that the time gaps between observations are large. Therefore,  an alternative suitable approach to modeling astronomical time series with large gaps between observations should be based on the solution of a difference equation of a discrete process. In this work we propose a novel model to fit irregular time series called the complex irregular autoregressive  (CIAR) model that is represented directly as a discrete-time process. We show that the model is weakly stationary and that it can be represented as a state-space system,  allowing efficient maximum likelihood estimation based on the Kalman recursions.  Furthermore, we show via Monte Carlo simulations that the finite sample performance of the parameter estimation is accurate. The proposed methodology is applied to light curves from periodic variable stars, illustrating how the model can be implemented  to detect poor adjustment of the harmonic model. This can occur when the period has not been accurately estimated or when the variable stars are multiperiodic. Last, we show how the CIAR model, through its state space representation, allows unobserved measurements to be forecast.
}


\maketitle
%%%%%%%%%%%%%%%%%%%%%%%%%%%%%%%%%%%%%%%%%%%%
\section{Introduction}
\label{sec:intro}

    Time-series theory is vast and provides a myriad of methods to model the serial correlation of a temporal sequence. Most of these methods assume that the observed values are measured regularly in time. However, there are many cases where this assumption is not valid. When this happens, we say that the time series is sampled irregularly.\\ 

An irregular time series is defined as a real valued sequence $\{y_{t_i}\}_{i=1}^n$ measured at observational times $t_1,\ldots,t_n$, such that the sequence $t_1,\ldots,t_N$ is strictly increasing and the distance between consecutive times $t_j-t_{j-1}$ is not constant.\\ 

Irregular time series can be observed in many disciplines. For example, natural disasters such as earthquakes do not occur regularly in time. Paleoclimate data are often sampled irregularly due to the difficulty in obtaining historical information. In astronomy, objects can be   observed irregularly in time because of, for example, the dependency of optical telescopes on clear skies.\\

Analyzing irregular time series is challenging because there are still few robust statistical tools available. In astronomy, some efforts have been made that attempt to model irregular time series. One approach is to transform the irregular times into regular ones by interpolation (\cite{Rehfeld_2011}). Other approaches implement  Gaussian processes to fit the light curves in the time domain (\cite{Foreman_2017}), but this solution can be computationally expensive. In some cases, regular time-series models have also been used to model astronomical data (for a review of such methods see e.g., \cite{Feigelson_2018}). Alternatively, the CARMA family of models has been  popular for fitting astronomical time series (\cite{Kelly_2014}), however these  models are defined as the solution of a differential stochastic equation, underlying the assumption of small time gaps between observations. Therefore, we consider it important to develop alternative models that can fit irregularly spaced time series under the assumption of discrete time gaps. \\ 

In \cite{Eyheramendy_2018} we introduced a new model called the
{ Irregular Autoregressive (IAR) model}  to fit unequally spaced time series. The IAR model is a discrete representation of the  continuous autoregressive model of order 1 (CAR(1)), which is strictly stationary and ergodic. The IAR has some extra flexibility when compared with the CAR(1).  In the model assumptions, the IAR can assume distributions other than Gaussian\footnote{We use the term Gaussian data to make reference to normally distributed time-series data.}. \\

The IAR model has two parameters that need to be estimated from the observed series. One parameter represents the variance of the process and the other one the value of the  autocorrelation function (ACF).  A limitation of the IAR model (and also of the CAR(1) model) is that the ACF derived from the estimated model parameters - via  eq.(1) - is positive only. This is a limitation that the regular autoregressive model does not have; the latter  can detect both positive and negative values of the ACF.\\ 

Negative values of the ACF appear often in some disciplines. For instance, they have been detected in some physical phenomena, such as the  velocity for hard spheres \citep[e.g.,][]{williams_2006,Alder_1970}. In financial time series, it is common to find significant negative autocorrelation for weekly and monthly stock returns (further discussion can be found in \cite{Sewell11}). Negative values of the  ACF are also generally observed in stocks with a high trading frequency  \citep[e.g.,][]{Conrad_1994,Campbell:1997}.\\ 

Well-known examples of time series with negative values for the ACF are the antipersistent processes, which are characterized as having negative values of the ACF for all positive lags (\cite{Bondon_07}).  One of the best known anti-persistent processes is the Kolmogorov energy spectrum of turbulence (\cite{Gao_2007}). There are several examples of antipersistent processes in meteorology. For instance, \cite{Ausloos_2001} detect antipersistence in the fluctuations of the Southern Oscillation Index, such as the sea level pressure. \cite{carvalho_2007} also detected antipersistence in temperature anomalies. Other examples are in financial data; for example, the electricity prices in some Canadian provinces show an antipersistent behavior \citep[e.g.,][]{Uritskaya_2015}.\\ 

The problem of detecting negative values of the ACF in irregularly sampled time series has scarcely been addressed in the literature. \cite{Chan_1987} showed that a discrete-time regular autoregressive model of order 1 (AR(1)) with negative coefficient can always  be embedded in a suitably chosen continuous-time ARMA(2,1) process, but this is not necessarily a parsimonious  solution. Alternatively, when an irregular time series has an antipersistent behavior it can be fitted by the  continuous-time ARFIMA (CARFIMA) process (\cite{Tsai_2009}) with an intermediate memory,that is, the Hurst parameter $H$ is such that $0 < H < 1/2$. In this work we propose another alternative that corresponds to an extension of the IAR model. We call this model the { Complex Irregular Autoregressive (CIAR)} model  because it needs complex numbers to represent the series.\\ 

Complex stochastic processes have been addressed by several authors, such as \cite{Miller_1974} and \cite{Picinbono_1997}. These processes are often used in signal-processing analyses since for example in telecommunication systems it is common to find complex signals. In addition, \cite{Dubois_1986} and \cite{Sekita_1991} introduced a complex extension of a regular autoregressive model, where they proposed to use the coefficients of these models for shape recognition. Furthermore, \cite{Martin_1999} proposed a method to estimate the parameters of the CAR(1) model in a complex time series. However, in these studies it is assumed that both the real and the imaginary part of the complex process are observed. The model proposed here assumes that only  the real part of the process is observed and the imaginary part is a latent process.\\ 

In this study, we apply the CIAR model to astronomical data, but the model can be applied in any other field in which irregular time series are available. In astronomy, the analysis of the temporal behavior of variable stars, transients, or supernovae attracts significant interest because  several properties of the objects can be obtained from the analysis of their light curves \citep[e.g.,][]{Richards_etal11, Kelly_2014, Elorrieta_etal16}. \\ 

In particular, in the analysis of variable stars the temporal modeling of the light curves is an important step to classify them. From a harmonic model fitted to the light curves of periodic variable stars,  several features  are extracted which describe the physical behavior of a specific class of variable stars.  We implement the CIAR model on the residuals of the harmonic model fitted on variable stars from OGLE \citep{OGLE} and Hipparcos \citep{hipparcos}. Under this scenario, our aim is to assess whether the harmonic model is capable of describing the temporal structure of the light curves of variable stars or  some autocorrelation remains on the residuals. \\ 

The structure of this paper is as follows. In Sect.~\ref{sec:methods} we present the irregular discrete time series models: the IAR model proposed by \cite{Eyheramendy_2018} and the CIAR model. In both cases the maximum likelihood estimation procedure is described. In the case of the CIAR model, we also provide expressions for forecasting unobserved measurements.  In Sect. \ref{sec:results} we assess the variance and bias of the estimated parameters of the CIAR model using Monte Carlo simulations. Furthermore, we perform simulations in order to compare the performance in fitting negative values of the  ACF of well-known time series models and the CIAR model.  In Sect.~~\ref{sec:Application} we apply the CIAR model to light curves of variable stars from the OGLE and Hipparcos surveys. We also implement the CIAR model on an AGN to demonstrate the forecasting procedure. We compare the results obtained with the IAR model and illustrate some applications for these data. The article ends with a  discussion and proposals for future work in Sect.~~\ref{sec:discussion}.\\ 

%%%%%%%%%%%%%%%%%%%%%%%%%%%%%%%%%%%%%
\section{Methods}
\label{sec:methods}

\subsection{Irregular Autoregressive (IAR) model} 
Letting $\{y_{t_j}\}$ be a time series observed at irregular times, where $\{t_j\}$ is an increasing sequence of observational times for $j=1,\ldots,n$, the irregular autoregressive (IAR) process is defined as 

\begin{equation}  \label{IAR} 
y_{t_j}=\phi^{t_j-t_{j-1}} \, y_{t_{j-1}} + \sigma_y \, \sqrt{1-\phi^{2(t_j-t_{j-1})}}  \, \varepsilon_{t_j}, 
\end{equation} 

\noindent
where $\phi$ is the parameter of the ACF and $\varepsilon_{t_j}$ is a white noise sequence\footnote{A white noise is an uncorrelated weakly stationary process.} with zero mean and unit variance. We note that 

\[\mathbb{E}(y_{t_j}) = 0$ and $\mathbb{V}(y_{t_j}) = \sigma^2_y ~~ \forall{t_j}. \]\\ 

The autocovariance function between two observational times, $s$ and $t$, with  $s<t,$ is  given by $\gamma(t-s)=E(y_t \, y_s)= \sigma_y^2 \,  \phi^{t-s}$, and the ACF, $\rho(t-s)=\frac{\gamma(t-s)}{\gamma(0)}= \phi^{t-s}$. Therefore, if $0 < \phi < 1$ the sequence $\{y_{t_j}\}$ corresponds to a second-order or weakly stationary process, that is, the time series has constant mean and finite second moment, and possesses an autocovariance function.\\ 

Furthermore, under some regularity conditions, the process $\{y_{t_j}\}$ is strictly stationary and ergodic (\cite{Eyheramendy_2018}).\\ 

We note that the IAR process is an extension of the regular autoregressive process. If $t_j-t_{j-1}=1$ is assumed, the IAR process becomes the autoregressive model of order $1$ (AR(1)). Also, the IAR process is equivalent to the continuous autoregressive process of order $1$ (CAR(1)) when a Gaussian distribution is assumed on the white noise sequence $\varepsilon_{t_j}$. However, the IAR process is  more flexible since it allows also nonGaussian data (\cite{Eyheramendy_2018}). \\

The finite past predictor of the process at time $t_j$ is given by,\\ 

\begin{equation} 
\widehat{y}_{t_j}=\phi^{t_j-t_{j-1}} \, y_{t_{j-1}}, \mbox{ for }j=2,\dots,n, 
\end{equation} 

\noindent
where the initial value is $\widehat{y}_{t_1}=0$. Consequently,  $e_{t_1}=y_{t_1}$ and $\nu_{t_1}=Var(e_{t_1})=\sigma_y^2$. Furthermore, $e_{t_j}=y_{t_j}-\widehat{y}_{t_j}$ is the innovation with variance $\nu_{t_j}=Var(e_{t_j})=\sigma_y^2 [1-\phi^{2(t_j-t_{j-1})}]$. \\ 

The estimation of the model parameters $\theta = (\sigma_y^2,\phi)$ can be performed by maximum likelihood. Minus  the log-likelihood of the process when a Gaussian distribution is assumed on $\varepsilon_{t_j}$ is given by 

\begin{equation}\label{eq:IARloglik} 
\ell(\theta)=\frac{n}{2}\log (2\pi)+\frac{1}{2}\sum_{j=1}^n \log \nu_{t_j} + \frac{1}{2}\sum_{j=1}^n \frac{e_{t_j}^2}{\nu_{t_j}}. 
\end{equation} 

\noindent
 For other distributional assumptions it can be written similarly. We can obtain the maximum likelihood estimator of $\sigma_y^2$ by  directly maximizing the log-likelihood \eqref{eq:IARloglik}, 

\begin{equation} 
\hat{\sigma_y}^2=\frac{1}{n}\sum_{j=1}^n\frac{(y_{t_j}-\widehat{y}_{t_j})^2}{\tau_{t_j}}, \mbox{ where }\tau_{t_j}=\nu_{t_j}/\sigma_y^2, 
\end{equation} 

\noindent
but it is not possible to find a closed form expression for the maximum likelihood estimator of $\phi$. However, iterative methods can be used.\\ 

A drawback of this model is that $\phi$ can only take values in the interval $[0,1]$, since a negative $\phi$ to the power of a real (and not integer) number is a complex number. Therefore, the IAR model only allows us to estimate positive values of the autocorrelation. To detect negative values of the ACF, the IAR model must be extended. In the following section we introduce the CIAR which allows for negative as well as positive autocorrelation to be modeled. 

\subsection{Complex Irregular Autoregressive (CIAR) model} 

To derive a complex extension of model \eqref{IAR}, we follow the approach of \cite{Sekita_1991}, which builds a complex autoregressive model for regular times. Supposing that $x_{t_j}$ is a complex valued sequence, such that, $x_{t_j} = y_{t_j}+ i z_{t_j}~\forall j=1,\ldots,n$,  and likewise, $\phi =  \phi^R + i \phi^I$ is the complex coefficient of the model and $\varepsilon_{t_j} = \varepsilon_{t_j}^R + i \varepsilon_{t_j}^I$ is a complex white noise, we define the CIAR process as, 

\begin{equation}  \label{CIAReq} 
y_{t_j}+ i z_{t_j}= (\phi^R + i \phi^I)^{t_j-t_{j-1}} \, (y_{t_{j-1}} + i z_{t_{j-1}}) + \sigma_{t_j}(\varepsilon_{t_j}^R + i \varepsilon_{t_j}^I), 
\end{equation} 

\noindent
where $\sigma_{t_j} = \sigma \, \sqrt{ 1-\left|\phi^{t_j-t_{j-1}}\right|^2}$ and $\left|.\right|$ is the modulus of a complex number. We assume that only the real part $y_{t_j}$ is observed and that the imaginary part $z_{t_j}$ is a latent process. In addition, $\varepsilon_{t_j}^R$ and $\varepsilon_{t_j}^I$, the real and imaginary part of $\varepsilon_{t_j}$, respectively, are assumed to be independent with zero mean and positive variance $\mathbb{V}(\varepsilon_{t}^R)=1$ and $\mathbb{V}(\varepsilon_{t}^I) = c,$ respectively, where $c$ is a fixed parameter that takes values in $\mathbb{R}^+$. Generally, we assume $c=1$, and the initial values are $y_{t_1} = \sigma \varepsilon_{t_1}^R$ and $z_{t_1} = \sigma \varepsilon_{t_1}^I$. In the following lemma we state some of the properties of this process.\\ 

\textbf{Lemma 1:} Consider the CIAR process $x_{t_j}$ described by Eq. \eqref{CIAReq}. Define $\gamma_0 =   \mathbb{E}(\overline{x}_{t_{j}}x_{t_j})$, $\gamma_k =   \mathbb{E}(\overline{x}_{t_{j+k}}x_{t_j}),$ and $\rho_k$ as the variance, autocovariance, and autocorrelation, respectively, of the process $x_{t_j}$. Subsequently,  the expected value, the variance, the autocovariance, and autocorrelation of the process  respectively satisfy  

\begin{enumerate}[a)] 
\item $\mathbb{E}(x_{t_j}) = 0,$ 
\item $\mathbb{V}(x_{t_j}) = \gamma_0 =   \mathbb{E}(\overline{x}_{t_{j}}x_{t_j}) = \sigma^2 (1+c),$\\ 
\item $\gamma_k =   \mathbb{E}(\overline{x}_{t_{j+k}}x_{t_j})=\overline{\phi^{\Delta_{k}}}\sigma^2 (1+c),$\\ 
\item $\rho_k =  \overline{\phi^{\Delta_{k}}}$,\\ 
\end{enumerate}

\noindent
where $\Delta_{k} = t_{j+k}-t_{j}$ denotes the time differences between the observational times $t_{j+k}$ and $t_{j}$. In addition,  $\overline{x}_{t_j}$ is the complex conjugate of $x_{t_j}$. See Appendix~\ref{sec:lem1} for a proof of this lemma.\\

If $|\phi| = |\phi^R + i \phi^I|< 1$, the results on Lemma $1$ above show that the complex sequence $x_{t_j}$ is a weakly stationary process. We note that the ACF $\rho_k$ of the CIAR process decays at  a rate $\phi^{t_{j+k}-t_j}$ (the so-called exponential decay). This autocorrelation structure is different to an antipersistent or intermediate memory CARFIMA process, since the ACF of the latter decays more slowly than an exponential decay. Thus, although both models can fit irregular time series with negative values of the ACF, the appropriate use of these models will depend on the correlation structure of the data. To make a decision about the most suitable model for the data, techniques based on the likelihood of the data, such as AIC, BIC, and so on, can be used.
 \\ 

In what follows, we  express the CIAR model in terms of  a state-space system. This representation  allows us: i) to implement the Kalman filter to obtain maximum likelihood estimators of the parameters of the model;{ and  ii) to forecast unobserved measurements.}\\ 

\subsection{State-space system} 

A linear state-space system may be described by the following equations. 

\begin{equation} \label{SSM} 
X_{t} = F_{t} X_{t-1}  + V_{t} 
,\end{equation} 
\begin{equation} \label{SSM1} 
Y_{t} = G X_t  + W_{t} 
,\end{equation} 

\noindent
where \eqref{SSM} is known as the state equation which determines a $v-$dimensional state variable $X_{t}$. Equation \eqref{SSM1} is called the observation equation, which expresses the $w-$dimensional observation $Y_{t}$. In addition, $F_t$ is a sequence of  ${v \times v}$ matrices called the transition matrices, and $G  \in \mathbb{R}^{w \times v}$ is the observation linear operator of the observation matrix. Finally, $W_t \sim WN(0,R_t)$, $V_t \sim WN(0,Q_t)$ and $V_t$ is uncorrelated with $W_t$.\\ 

In order to represent the CIAR model in a state-space system we need to rewrite equation \eqref{CIAReq}. In the following lemma an alternative way of writing the CIAR model is proposed.\\ 

\textbf{Lemma 2:} The CIAR process described by Eq. \eqref{CIAReq} can be expressed by the following equation.

\begin{equation}  \label{CIAR1}
y_{t_j}+ i z_{t_j}= (\alpha_{t_j}^{R} + i \alpha_{t_j}^{I}) \, (y_{t_{j-1}} + i z_{t_{j-1}}) + \sigma_{t_j}(\varepsilon_{t_j}^R + i \varepsilon_{t_j}^I),\\
\end{equation}

\noindent
where\\
\noindent
 $\alpha_{t_j}^{R} = |\phi|^{\delta_{j}} \cos(\delta_{j} \psi)$, $\alpha_{t_j}^{I} = |\phi|^{\delta_{j}} \sin(\delta_{j} \psi)$, $\delta_{j} = t_j-t_{j-1}$, $\psi = \arccos\left( \frac{\phi^{R}}{|\phi|}\right)$ and $\phi = \phi^R + i \phi^I$. See Appendix~\ref{sec:lem2} for a proof of this lemma.\\

By following the representation of the CIAR model described in Lemma 2, we can express the observed process as

\begin{equation}
y_{t_j} = \alpha_{t_j}^{R}y_{t_{j-1}} -  \alpha_{t_j}^{I}z_{t_{j-1}} + \sigma_{t_j}\varepsilon_{t_j}^R
,\end{equation}

\noindent
 and the latent process as

\begin{equation}
z_{t_j} = \alpha_{t_j}^{I}y_{t_{j-1}} +  \alpha_{t_j}^{R}z_{t_{j-1}} + \sigma_{t_j}\varepsilon_{t_j}^I.
\end{equation}

 We note that the observed process $y_{t_j}$ is an IAR with parameter $\phi$, if we assume $\alpha_{t_j}^{I} = 0$ and $\alpha_{t_j}^{R} = \phi^{t_j-t_{j-1}}$. In addition, it is straightforward to show that $\alpha_{t_j}^{I} = 0$ is equivalent to $\phi_I=0$.\\

Another important consideration is that the observed process $y_{t_j}$ does not depend directly on $\varepsilon_{t_j}^I$. Consequently, the variance of the imaginary part $c$ is a nuisance parameter, in the sense that it takes any value in $\mathbb{R}^+$ and does not cause significant changes in the model.\\

Equation \eqref{CIAR1} can be represented by the state-space system of Eq. \eqref{SSM} - \eqref{SSM1} assuming $t=t_j$ and $X_{t_j} = \left(\begin{array}{c} y_{t_j} \\  z_{t_j} \end{array} \right)$. Given that only  $y_{t_j}$ are actually observed, we obtain  $Y_t = y_{t_j}$. Therefore, $G=\left(\begin{array}{cc} 1 & 0 \end{array} \right)$ is the observation matrix under this representation.\\ 

To complete the specification we define the transition matrix as $F_{t_j} = \left(\begin{array}{cc} \alpha_{t_j}^{R} & -\alpha_{t_j}^{I} \\ \alpha_{t_j}^{I} & \alpha_{t_j}^{R} \end{array} \right)$ and the noise of equations \eqref{SSM} and \eqref{SSM1} as $V_{t_j} = \sigma_{t_j} \left(\begin{array}{c} \varepsilon_{t_j}^R \\ \varepsilon_{t_j}^I \end{array} \right)$ and $W_{t_j} = 0,$ respectively. Finally, the observation and state equations of the state-space representation of the CIAR model are,

\begin{equation}  \label{CIARSS}
\left(\begin{array}{c} y_{t_j} \\  z_{t_j} \end{array} \right)= \left(\begin{array}{cc} \alpha_{t_j}^{R}  & -\alpha_{t_j}^{I} \\ \alpha_{t_j}^{I} & \alpha_{t_j}^{R} \end{array} \right)\left(\begin{array}{c} y_{t_{j-1}} \\  z_{t_{j-1}} \end{array} \right) + \sigma_{t_j} \left(\begin{array}{c} \varepsilon_{t_j}^R \\ \varepsilon_{t_j}^I \end{array} \right)
,\end{equation}
\begin{equation}  \label{CIARSS2}
y_{t_j} = \left(\begin{array}{cc} 1 & 0 \end{array} \right) \left(\begin{array}{c} y_{t_j} \\  z_{t_j} \end{array} \right)
.\end{equation}

We note that in this representation, the transition matrix and the variance of noise term of the state equation $Q_{t_j} =|\sigma_{t_j}|^2 \mathbb{V}\left(\varepsilon_{t_j} \right)$ depend on time.\\ 

\textbf{Lemma 3:} Now we let $\alpha_{t_j} = \alpha_{t_j}^{R} + i \alpha_{t_j}^{I}$. If $\sup|\alpha_{t_j}|  < 1$, the process in Eqs. \eqref{CIARSS}-\eqref{CIARSS2} is stable and has a unique stationary solution given by

\begin{equation} \label{eq:lemma3}
$$X_{t_j} = V_{t_j} + \mathop{\sum}_{k=1}^{\infty} V_{t_{j-k}} \mathop{\prod}_{i=0}^{k-1} F_{t_{j-i}}$$
\end{equation}

\noindent
where $V_{t_{j-k}} = \sigma_{t_{j-k}} \left(\begin{array}{c}  \varepsilon_{t_{j-k}}^R \\  \varepsilon_{t_{j-k}}^I \end{array} \right)$. Appendix~\ref{sec:lem3} presents a proof of this lemma.\\

%%%%%%%%%%%%%%%%%%%%%%%%%%%%%%%%%%%%%%%%%%%%
\subsection{Estimation} 

For the state-space model  of Eq. \eqref{SSM} - \eqref{SSM1}, the one-step predictors $\hat{X}_{t_j} = P_{t_{j-1}}(X_{t_j})$ and their error covariance matrices $\Omega_{t_j} = \mathbb{E}[(X_{t_j} - \hat{X}_{t_j})(X_{t_j} - \hat{X}_{t_j})']$ are unique and determined by the initial values: $\hat{X}_{t_1} = \left(\begin{array}{c} 0 \\ 0 \end{array} \right)$ and $\Omega_{t_1} = \mathbb{E}[(X_{t_1} - \hat{X}_{t_1})(X_{t_1} - \hat{X}_{t_1})']$. Using the properties of  model \eqref{CIAReq} we can rewrite $\Omega_{t_1}$ as, 

$$\Omega_{t_1}  =  \sigma^2\left(\begin{array}{cc} 1   & 0\\ 0 &  c   \end{array} \right).$$ 

The Kalman recursions, for $j=1, \ldots,n-1$ are defined as follows.

\begin{equation} 
\centering
\begin{split}
\Lambda_{t_j} &= G_{t_j} \Omega_{t_j} G_{t_j}'  ,\\
\Theta_{t_j} &= F_{t_j}\Omega_{t_j} G_{t_j}' ,\\
\Omega_{t_{j+1}} &= F_{t_j} \Omega_{t_j} F_{t_j}' + Q_{t_j}  - \Theta_{t_j} \Lambda_{t_j}^{-1} \Theta_{t_j}' ,\\
\nu_{t_j}  &= y_{t_j} - G_{t_j} \hat{X}_{t_j},\\
\hat{X}_{t_{j+1}}  &= F_{t_j} \hat{X}_{t_j} +  \Theta_{t_j} \Lambda_{t_j}^{-1} \nu_{t_j},
\end{split}
\label{eq:Kalman}
\end{equation}

\noindent
where $\{\nu_{t_j}\}$ is called the innovation sequence.\\ 
The maximum likelihood estimators of the CIAR model parameters $\phi^R$ and $\phi^I$  can be obtained by minimizing the reduced likelihood defined as, 
$$\ell(\phi) \propto  \frac{1}{n} \mathop{\sum}\limits_{j=1}^n \left( \log (\Lambda_{t_j}) + \frac{\nu_{t_j}^2}{\Lambda_{t_j}} \right),$$ 
where $\Lambda_{t_j}$ and $\nu_{t_j}$ come from the Kalman recursion. We developed scripts in the statistical language/software R and in Python to perform the estimation of the parameters of the model using Kalman recursions.\\ 

Another feature of this model is that it allows forecasting, the process of which is explained below.\\

\subsection{Forecasting} 
 
 Using the space-state equations \eqref{CIARSS} - \eqref{CIARSS2} it is straightforward to define a forecasting expression for the CIAR model. We note that from the Kalman filter recursions \eqref{eq:Kalman} we obtain the estimates of the state $\hat{X}_{t_{j}}$, $\Lambda_{t_j}$, $\Theta_{t_j}$ and $\nu_{t_j}$ at the last time point $t_n$. A one-step forecasting for the CIAR model can be obtained from the following equations.

\begin{equation} 
\centering
\begin{split}
\hat{X}_{t_{n+1}} &= F_{t_{n+1}} \hat{X}_{t_n} + \Theta_{t_n} \Lambda_{t_n}^{-1} \nu_{t_n} \\
\hat{Y}_{t_{n+1}} &= G_{t_{n+1}} \hat{X}_{t_{n+1}}\\
 \end{split}
\label{eq:forecasting}
,\end{equation}

\noindent
where $F_{t_{n+1}} =  |\phi|^{\delta} \left(\begin{array}{cc} \cos(\delta \psi)  & -\sin(\delta \psi) \\ \sin(\delta \psi) & \cos(\delta \psi)  \end{array} \right)$ and $\delta=t_{n+1} - t_{n}$.\\

In addition, a confidence interval for the forecast $\hat{Y}_{t_{n+1}}$ can be defined by,

\begin{equation} 
\hat{Y}_{t_{n+1}} \pm z_{1-\alpha/2} \sqrt{\mathbb{V}(e_{t_{n+1}})}
\label{eq:confidence}
,\end{equation}

\noindent
where $e_{t_{n+1}} = Y_{t_{n+1}} - \hat{Y}_{t_{n+1}}$ is the forecast error with variance $\mathbb{V}(e_{t_{n+1}}) = \sigma^2(1-\left|\phi^{t_j-t_{j-1}}\right|^2)$ and $z_{1-\alpha/2}$ is the $1-\alpha/2$ quantile of the standard normal distribution.\\

%%%%%%%%%%%%%%%%%%%%%%%%%%%%%%%%%%%%%%%%%%%%
\section{Simulation results}
\label{sec:results}

\subsection{Assessing the estimation performance of the CIAR model} 
\label{ssec:Montecarlo}

In this section, we assess the performance of the estimation procedure proposed for the CIAR model. We perform Monte Carlo experiments based on 1000 repetitions of each simulation. In each repetition we generate a CIAR sequence from the model \eqref{CIAReq} using coefficients with different positive and negative values for the real part  $\phi^R$. In addition, both the imaginary part of the coefficient and the imaginary variance are set to $\phi^I = 0$ and $c=1$,  and the real and imaginary errors are assumed to follow a Gaussian distribution with a mean equal to zero and a variance of one. The irregular times are generated using the following mixture of two exponential distributions, 

\begin{equation} 
f(t|\lambda_1,\lambda_2,\omega_1,\omega_2)=\omega_1g(t|\lambda_1)+\omega_2g(t|\lambda_2).
\label{mixture} 
\end{equation}

We choose $\lambda_1=15$ and $\lambda_2=2$ as the means of each exponential distribution, respectively, and $\omega_1=0.15$ and $\omega_1=0.85$ as their respective weights. Under these parameters, the mean of the time differences in the simulated data is $\approx 3.95$.  The rationale for choosing this distribution came from mimicking time gaps observed in surveys such as the VVV (\texttt{https://vvvsurvey.org/}) where some observations are very close to each other followed  by observations that are more separated from each other.\\ 

Table \ref{tab:sim2}  shows the results of the Monte Carlo simulations, which suggest that the finite-sample performance of the proposed methodology is accurate, both for positive and negative values of the parameter 
$\phi^R$. In addition, we also estimate the parameter $\phi$ of the IAR model in each simulation. A precise IAR coefficient estimate is obtained when $\phi^R$ is positive, but when the CIAR process is generated with negative $\phi^R$ the estimation of the IAR coefficient is close to zero. This result is  important since it shows that the IAR model cannot detect negative values of the ACF, unlike the CIAR model. Finally, we note that the accuracy of the estimated values does not depend on the magnitude of the coefficient. 

Another method commonly used for estimating autocorrelation in irregular time series is the discrete correlation function (DCF;\citep{Edelson}). This method is based on computing a Pearson correlation coefficient between data pairs. Each data pair is composed by observations with a time difference in the interval $(\tau - \delta/2,\tau + \delta/2)$ where $\tau$ is the order of the autocorrelation and $\delta$ is the bin window size. In order to assess whether this method can detect the autocorrelation of the CIAR process, we implement the DCF using the package {\em sour} of R \citep{Edelson_2017}. The last two columns in Table \ref{tab:sim2} correspond to the mean of the DCF estimates of order $\tau=1$ and its standard deviation, respectively. We note that the DCF estimates are close to the $\phi^R$ parameter when it is positive, but with a large standard deviation. On the other hand, when $\phi^R$ is negative, the DCF estimates are not accurate.\\ 

\begin{table*} 
\centering 
\caption{\em Maximum likelihood estimation of complex $\phi$ computed by the CIAR model in the real part of the simulated CIAR data. The observational times are generated using a mixture of exponential distribution with $\lambda_1=15$ and $\lambda_2=2$, $\omega_1=0.15$ and $\omega_2=0.85$. \label{tab:sim2}} 
\begin{tabular}{rr|rrr|rrr|rr|rr} 
  \hline 
Case & N & ${\phi}^R$ & $\widehat{\phi}^R$ & SD($\widehat{\phi}^R)$  & ${\phi}^I$ & $\widehat{\phi}^I$ & SD($\widehat{\phi}^I)$ & $\widehat{\phi}_{\mbox{IAR}}$ & SD($\widehat{\phi}_{\mbox{IAR}}$) & DCF & SD(DCF)\\ 
   \hline 
  1 & 300 & 0.999 & 0.9949 & 0.0036 & 0& 0.0009 & 0.0030 & 0.9949 & 0.0036 & 0.9743 & 0.1273 \\
  2 & 300 & 0.9 & 0.8960 & 0.0187 & 0& 0.0116 & 0.0413 & 0.8950 & 0.0188 & 0.8809 & 0.1315 \\
  3 & 300 & 0.7 & 0.6967 & 0.0412 & 0& 0.0557 & 0.0819 & 0.6948 & 0.0406 & 0.6909 & 0.1279 \\ 
  4 & 300 & 0.5 & 0.4942 & 0.0596 & 0& 0.0849 & 0.1111 & 0.4965 & 0.0569 & 0.5004 & 0.1240 \\ \hline 
  5 & 300 & -0.999 & -0.9984 & 0.0012 & 0& 0.0001 & 0.0009 & 0.0626 & 0.0265 & -0.6260 & 0.1038 \\
  6 & 300 & -0.9& -0.8991 & 0.0154 & 0& 0.0014 & 0.0134 & 0.0643 & 0.0299 & -0.5644 & 0.1166 \\
  7 & 300 & -0.7 & -0.6991 & 0.0414 & 0& 0.0061 & 0.0354 & 0.0628 & 0.0289 & -0.4382 & 0.1114 \\
  8 & 300 & -0.5 & -0.4971 & 0.0717 & 0& 0.0091 & 0.0607 &  0.0589 & 0.0283 & -0.3152 & 0.1089 \\ \hline 
\end{tabular} 
\end{table*}

\subsection{Comparison of the CIAR with other time series models} 
\label{ssec:Comparing} 

\begin{center} 
\begin{figure*} 
\begin{minipage}{0.47\linewidth} 
\includegraphics[width=\textwidth]{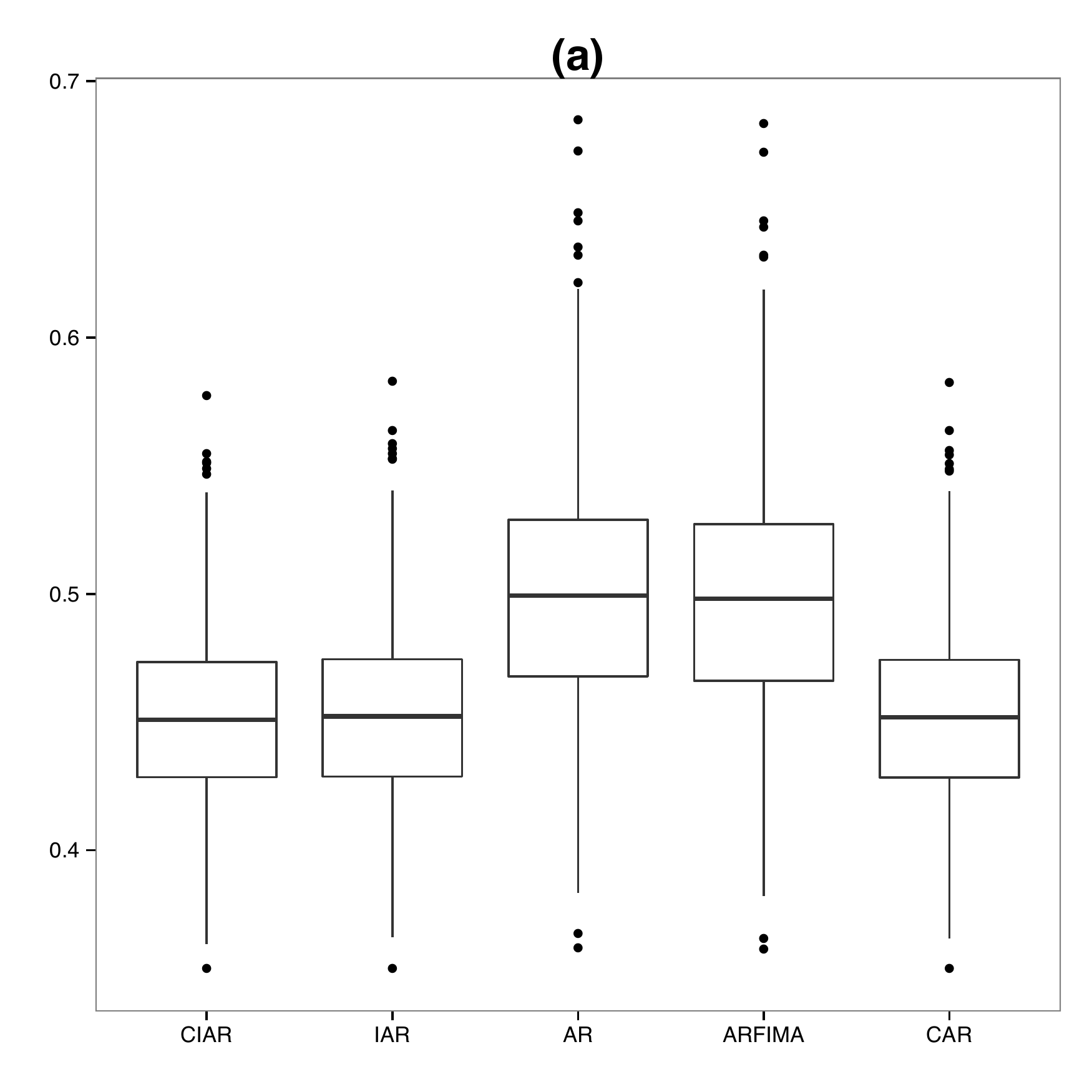} 
\end{minipage} 
\begin{minipage}{0.47\linewidth} 
\includegraphics[width=\textwidth]{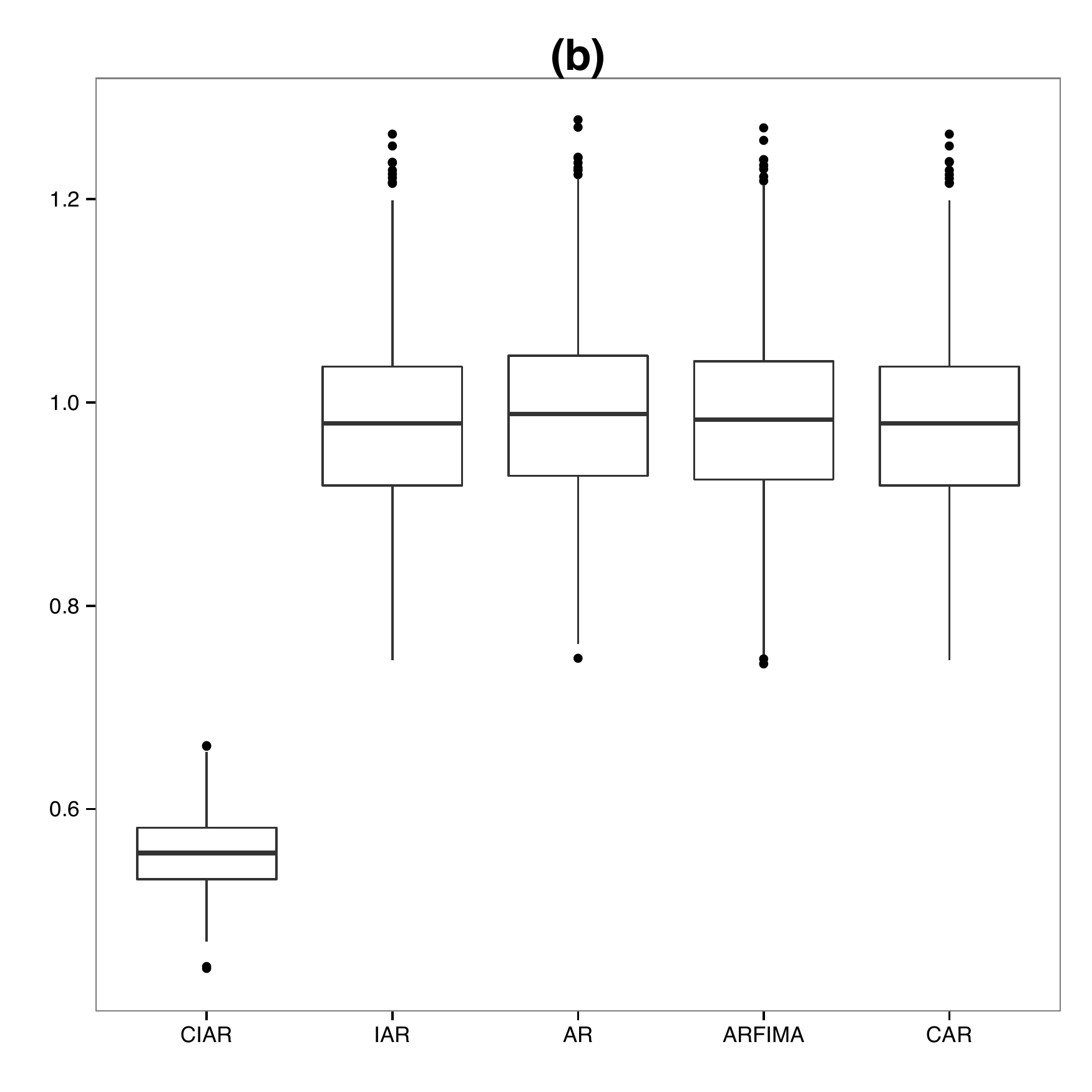} 
\end{minipage} 
\caption{Boxplot of the root mean squared error computed for the fitted models on the 1000 sequences simulated of the real part of the CIAR process. In a) each CIAR process was generated using $\phi^R = 0.99$. In b) each CIAR process was generated using $\phi^R = -0.99$. The other parameters of the models are defined as $\phi^I = 0$, $c=0$ and length $n=300$. The observational times are generated using a mixture of Exponential distribution with $\lambda_1=15$ and $\lambda_2=2$, $\omega_1=0.15$ and $\omega_2=0.85$ .\label{fig:RMSECIAR}} 
\end{figure*} 
\end{center}

We perform a simulation experiment to compare the performance of well-known time series models, that is, IAR, AR(1), ARFIMA and CAR(1), in fitting a CIAR process. This experiment  consists in generating 1000 sequences of the CIAR process $\{y_1,\ldots,y_n\}$ with length $n=300$. In order to generate positive and negative values of the ACF, $\phi^R = 0.99$ and $\phi^R = -0.99$ are used. The remaining parameters are set as $\phi^I = 0$, $c=1$ and the irregular times are generated with a mixture of two exponential distributions. We then fit each sequence using the different  time-series models, and the CIAR model. The AR(1), ARFIMA, and CAR(1) models were estimated by maximum likelihood using the functions \texttt{arima}, \texttt{arfima}, and \texttt{cts}, respectively, from the statistical software R. To compare its performance we compute the root mean squared error (RMSE). Figure~\ref{fig:RMSECIAR} a) shows that the RMSE estimated by the irregular time-series models is smaller in comparison with that of  either the AR or ARFIMA model which assume regular sampling. However, as can be seen in Figure~\ref{fig:RMSECIAR}b the CIAR model shows significantly better performance in fitting the negatively correlated processes than the other implemented models. Therefore, we verify that a CIAR process with a large and negative value of $\phi^R$ cannot be correctly modeled with the conventional time-series models, including those that assume irregular sampling. \\ 

\subsection{Computation of  the autocorrelation in a harmonic model} 
\label{ssec:Harmonic} 

An important advantage of the CIAR process over other time-series models for unequally spaced observation is its ability to model  weakly stationary time series with negative as well as positive values of the ACF.  A well-known example of a time series that can be negatively correlated is the following harmonic process,\\ 

\begin{equation} 
y_{t_i} = A \sin (f t_i + \psi) + \epsilon_{t_i} 
\label{harmonic} 
,\end{equation} 

\noindent
where $f$ is the frequency of the process and $\epsilon_{t_i}$ is a Gaussian sequence with a mean of zero and a variance of  $\sigma^2$. In addition, the amplitude $A$ is a fixed parameter and the phase $\psi$ is a random variable with uniform distribution between $-\pi$ and $\pi$. We note that the weakly stationarity of the process $y_{t_i}$ is guaranteed when this distribution for $\psi$ is assumed (for more details, see e.g., \cite{Lindgren:2013}). Assuming regular observational times, the one-step autocorrelation is given by $\rho_1 = \cos(f)$ (\cite{Broersen:2006}). This result is also satisfied under irregular times. We note that this autocorrelation is negative for $f \in (\pi/2,\pi)$. In addition, we note that for higher-frequency values the harmonic process (\ref{harmonic}) becomes more anti-persistent (\cite{Alperovich_2017}). \\ 

\begin{figure}
\centering 
\includegraphics[scale=0.45]{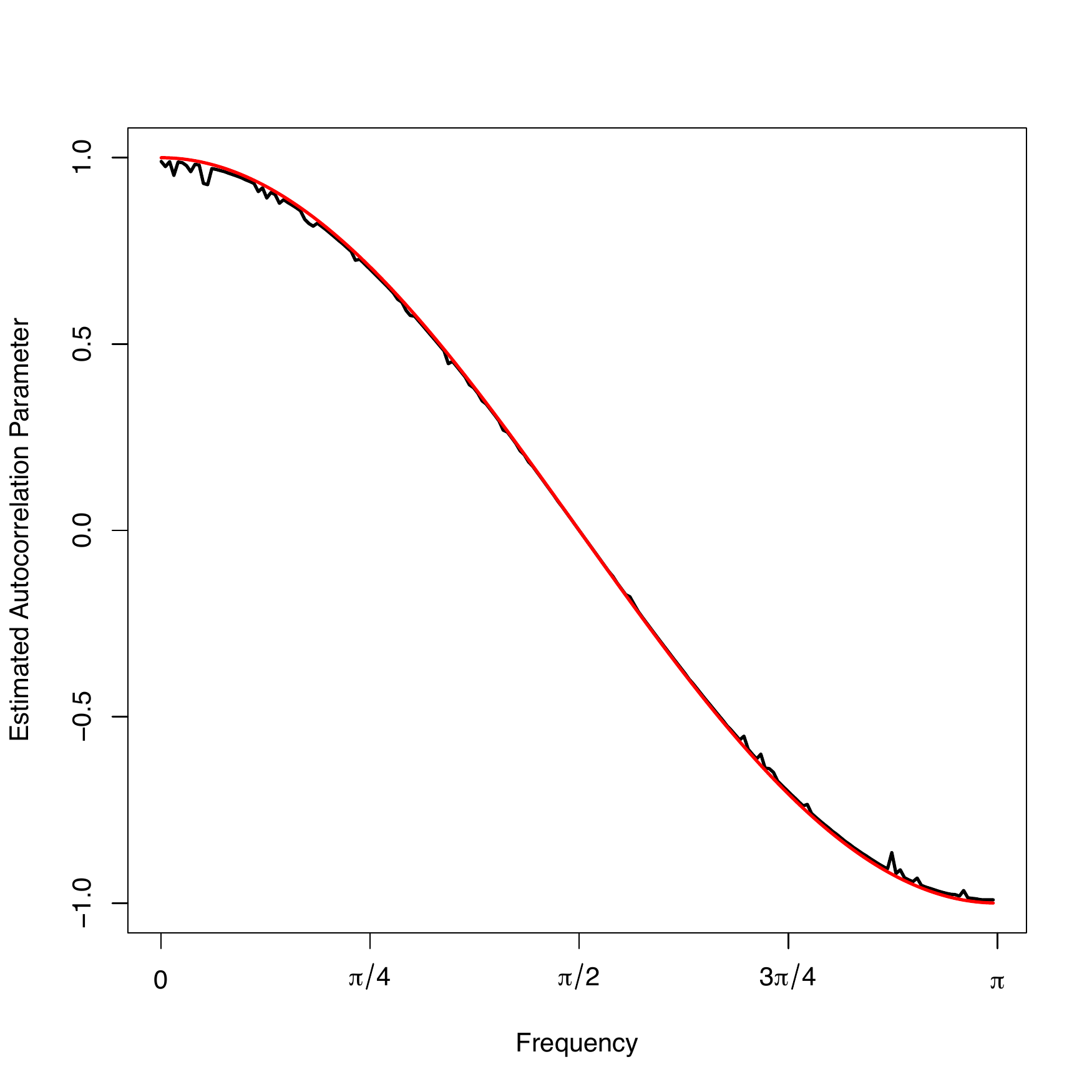} 
\caption{Estimated coefficients $\phi^R$ (y-axis) by the CIAR  model in $k=200$ harmonic processes generated using frequencies (x-axis) in the interval $(0,\pi)$. The black line corresponds to the coefficients estimated by the CIAR model.  The red line is the theoretical autocorrelation of the process $y_{t_i}$ \label{fig:Harmonic}} 
\end{figure}

We performed a simulation study in order to assess whether the CIAR model can detect the correlation structure of an irregular harmonic model. We generated the irregular observational times ${t_i}$ with $i=1,\ldots,n$ using the mixture of exponentials distributions (Eq. \eqref{mixture}). The process $y_{t_i}$ is simulated with length $n=300$, amplitude $A = 20,$ and a unit variance for $\epsilon_{t_i}$. Simulations using \eqref{harmonic} are run with $k=200$ different frequencies taken equally spaced from the interval $(0,\pi)$. We fit a CIAR  model to each simulated sequence. In Figure \ref{fig:Harmonic} the parameters estimated (y-axis) from the  CIAR model for each of the $200$ frequencies (x-axis) are shown as the black line, and the mapping of the frequency values to the theoretical autocorrelation $\cos(f)$ is depicted as a red line. We note that the $\phi^R$ parameter estimated by the CIAR model fits the theoretical values almost perfectly.

%%%%%%%%%%%%%%%%%%%%%%%%%%%%%%%%%%%%%%%%%%%%

\section{Application of the CIAR model to astronomical data} 
\label{sec:Application} 

%\vspace{0.1in}
In this section we show some results of the implementation of the CIAR model in astronomical time series. Astronomical data are naturally irregularly sampled since it is not always possible to get observational data from optical telescopes due to a dependency on clear skies. Many astronomical objects, such as variable stars, transients, and supernovae, can be characterized by their brightness. Generally, this brightness is represented by a time series (light curve) of this object. We apply the CIAR model to the residuals of a harmonic model fitted to  the light curves of variable stars.\\ 

\subsection{Modeling light curves} 
\label{ssec:OGLE} 

One of the most important challenges in the analysis of variable stars is to classify them based on their temporal behavior. The pulsating variables represent  a particular class of variable stars that are characterized by having a periodical behavior. Therefore, the light curves of pulsating variable stars are generally fitted by a harmonic model (see e.g.\citep[e.g.]{Debosscher_etal07,Richards_etal11,Elorrieta_etal16}).The p-harmonic model is defined as 

\begin{equation} 
y(t) = \beta_0 + \mathop{\sum}\limits_{j=1}^p (\alpha_{1j} sin(2\pi f_1 jt) +  \beta_{1j} cos(2\pi f_1 jt)) + \epsilon(t) 
\label{pArm} 
,\end{equation} 

\noindent
 where $f_1$ corresponds to the dominant frequency estimated by the generalized Lomb-Scargle periodogram (GLS; \cite{Zechmeister_etal09}). Following \cite{Debosscher_etal07},for example, $p=4$ is assumed to fit the light curves. The main goal of implementing  the irregular time series models in light curves from periodic variable stars is to detect whether the harmonic model is sufficient to capture all the temporal dependency in the light curve. Otherwise, the residuals of the harmonic model remain with autocorrelation.\\ 
  
We apply the CIAR model to the residuals of the harmonic model fitted on light curves of variable stars from the optical surveys OGLE and Hipparcos. We use these data since many classes of variable stars are available in the catalogue of these surveys. %In addition, these surveys are observed in the optical I-band, and therefore the brightness magnitude of a star has smaller errors. We  compare the results obtained by the IAR and CIAR models. 

\begin{figure}
\centering 
\includegraphics[scale=0.45]{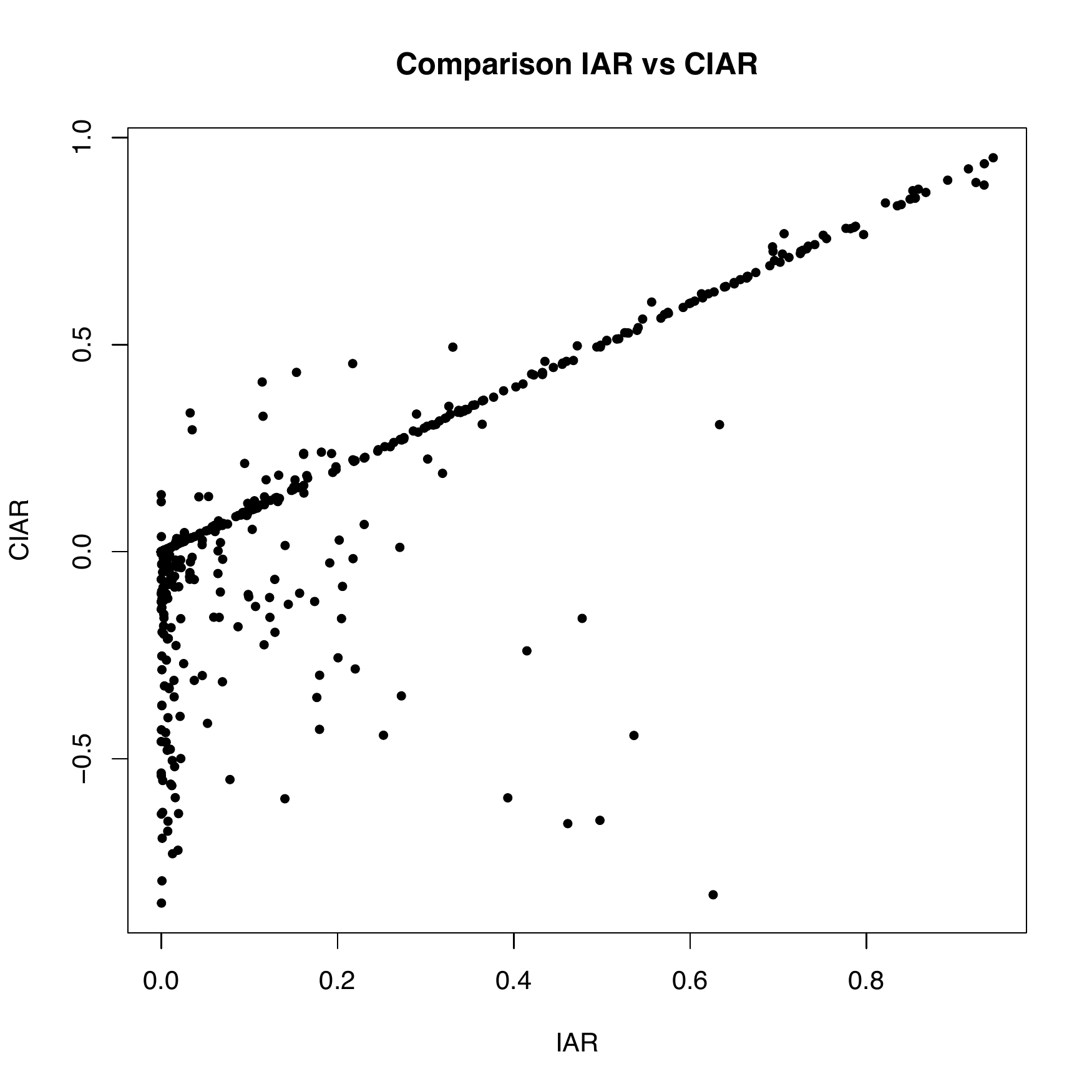} 
\caption{Plot of  $\phi^R$ vs. $\phi_{IAR}$, the parameter coefficients estimated by the CIAR and IAR models, respectively,  from light curves in OGLE and HIPPARCOS.\label{CIAR}} 
\end{figure}

In order to fit both models on these data, each light curve of the OGLE and Hipparcos surveys was fitted by a four-harmonic model (equation \eqref{pArm}). The residuals of each harmonic model are then fitted using both the IAR and CIAR models. In Figure \ref{CIAR}, we note that there is a high correlation between the estimated coefficients $\hat{\phi}_{IAR}$ and $\hat{\phi^R}$ of the IAR and CIAR models, respectively, on the light curves when both values are positive, which is consistent with the results of the Monte Carlo simulations.\\ 

However, we observe that several cases identified as uncorrelated  by the IAR model, that is, $\phi_{IAR}\approx 0$, have a high but negative autocorrelation estimated by the CIAR model. In other words, these light curves remain with negative dependency structure on the residuals after fitting the harmonic model.  Double-mode cepheids (DMCEP) and eclipsing binaries (EB) make up the majority of these variable stars.\\ 

The time structure detected can be due to the fact that a light curve was incorrectly fitted by the harmonic model, for example when using an incorrect period. Also the CIAR model can detect a correlation structure in the residuals of a harmonic model fitted to the light curve of a multiperiodic variable star. Intuitively, if the light curve has two or more periods, a harmonic model fitted using only the dominant frequency is not sufficient to account for all its temporal dependency structure.\\ 

\begin{figure}
\centering 
\includegraphics[scale=0.45]{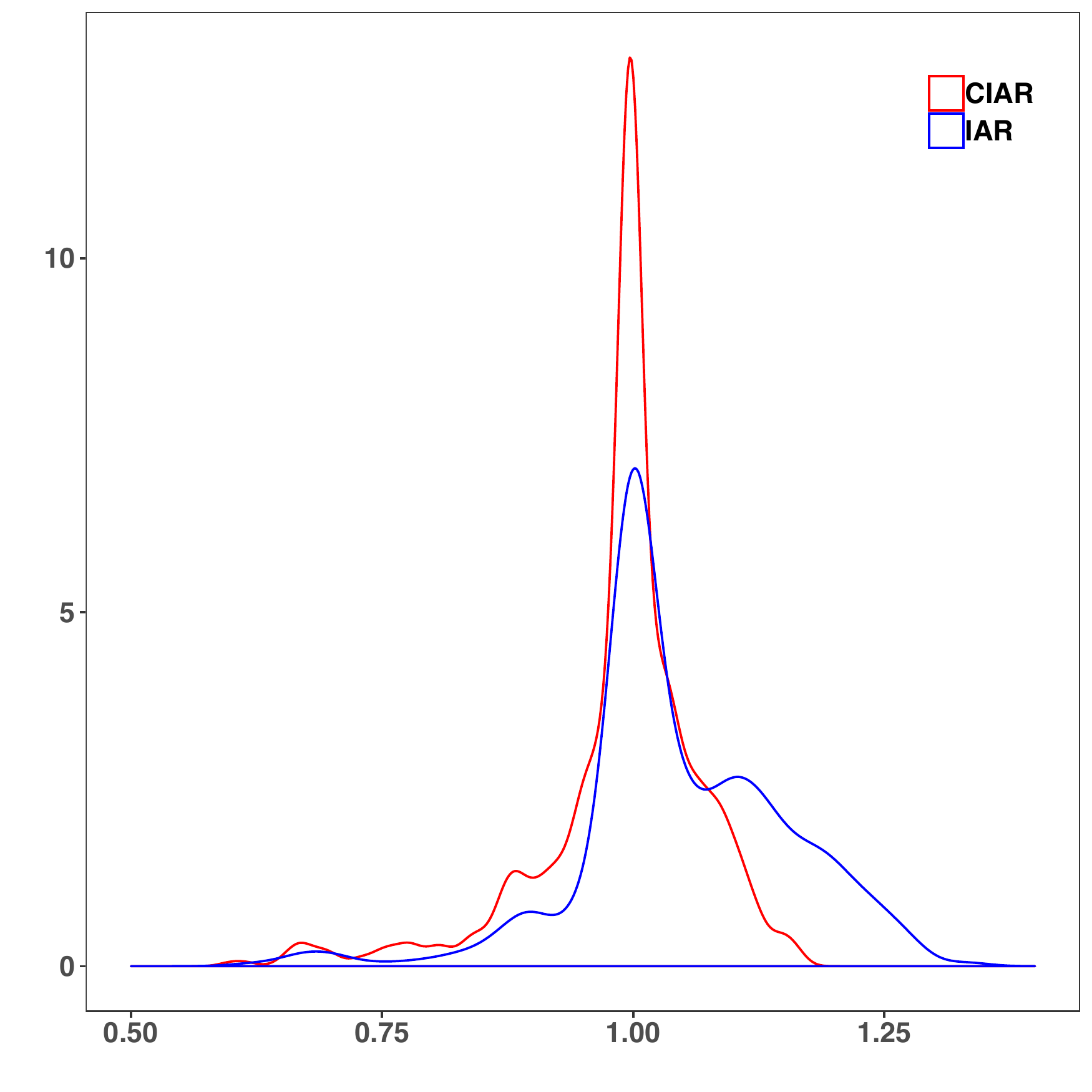} 
\caption{Kernel Density of the RMSE computed for the residuals of harmonic fit in the light curves when the CIAR coefficient is negative. The red density corresponds to the RMSE computed using the CIAR model, and the blue density corresponds to the RMSE computed using the IAR model.\label{RMSEOH}} 
\end{figure}

It is also important to assess the goodness of fit performance of the IAR and CIAR models on these data. We compute the RMSE after fitting each irregular model on the residuals of the harmonic model. These results do not vary significantly when $\phi^R$ is positive. However, if this coefficient is negative, the RMSE estimated by the CIAR model is smaller than the one obtained when we fit these data using the IAR model, as can be seen in Figure \ref{RMSEOH}. This result is consistent with those obtained in Section \ref{ssec:Comparing}. \\ 

\subsection{Statistical test for the autocorrelation parameter} 
\label{ssec:Test} 

The magnitude of the coefficients estimated by the irregular time series models depends on the period of the light curve. Light curves with short periods have, in general,  smaller estimated coefficients. In other words, a small value of the estimated coefficient does not always imply uncorrelated residuals. If the light curve has a large period, a small coefficient could  mean uncorrelated residuals. However, if the variable star has a shorter period, it could  mean the opposite. Therefore, we cannot make decisions based only on the value of the parameter estimated by the model. \\

To overcome this problem we designed a statistical test that can assess when a value of $\phi$ is significantly different  from (larger than) the values that can arise by chance. To do that, we estimated the distribution of the parameter $\widehat{\phi}^R$ for time series with autocorrelation. We then assessd how likely it is to observe a particular value of  $\widehat{\phi}^R$ under this distribution. Following the methodology proposed by \cite{Eyheramendy_2018} we selected 38 different frequencies to fit each light curve incorrectly, where each of them is a variation of the correct period in the interval $(f_1-0.5f_1,f_1+0.5f_1)$. The factor $f_1$ corresponds to the correct frequency of the light curve. To develop the test we assumed that the $\log(|\widehat{\phi}^R|)$ follows a Gaussian distribution, estimated using the  $38$ $\widehat{\phi}^R_i, i=1,\ldots,38$  obtained from the CIAR fit of the time series that each assume a different incorrect frequency from the interval. Consequently, the null hypothesis here is that the $\log(|\widehat{\phi}^R|)$ belongs to this Gaussian distribution. We note that the test is formulated so that each estimated coefficient is compared only with the estimated ones in the same light curve fitted incorrectly. 

\begin{center} 
\begin{figure*}
\centering 
\begin{minipage}{0.32\linewidth} 
\includegraphics[width=\textwidth]{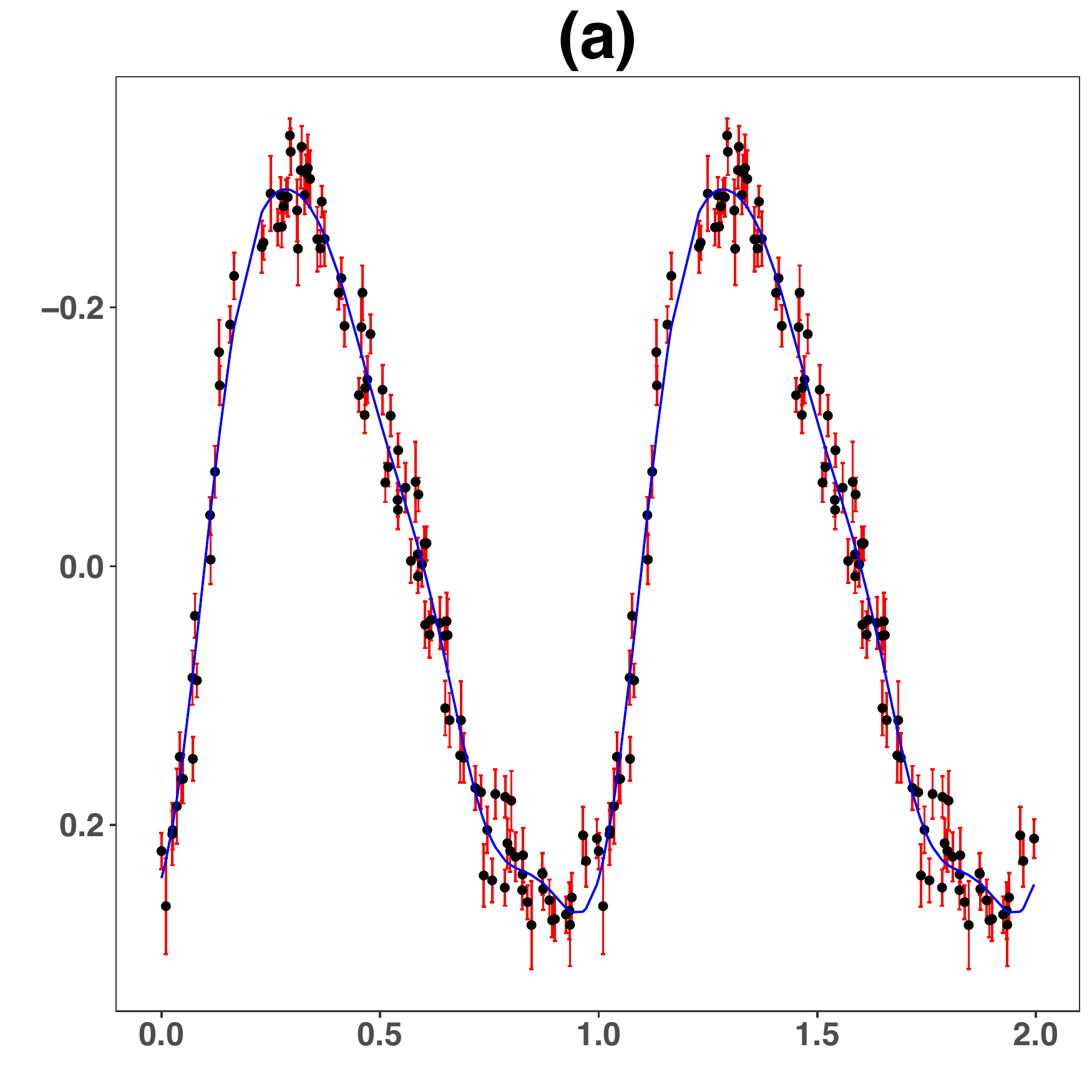} 
\end{minipage} 
\begin{minipage}{0.32\linewidth} 
\includegraphics[width=\textwidth]{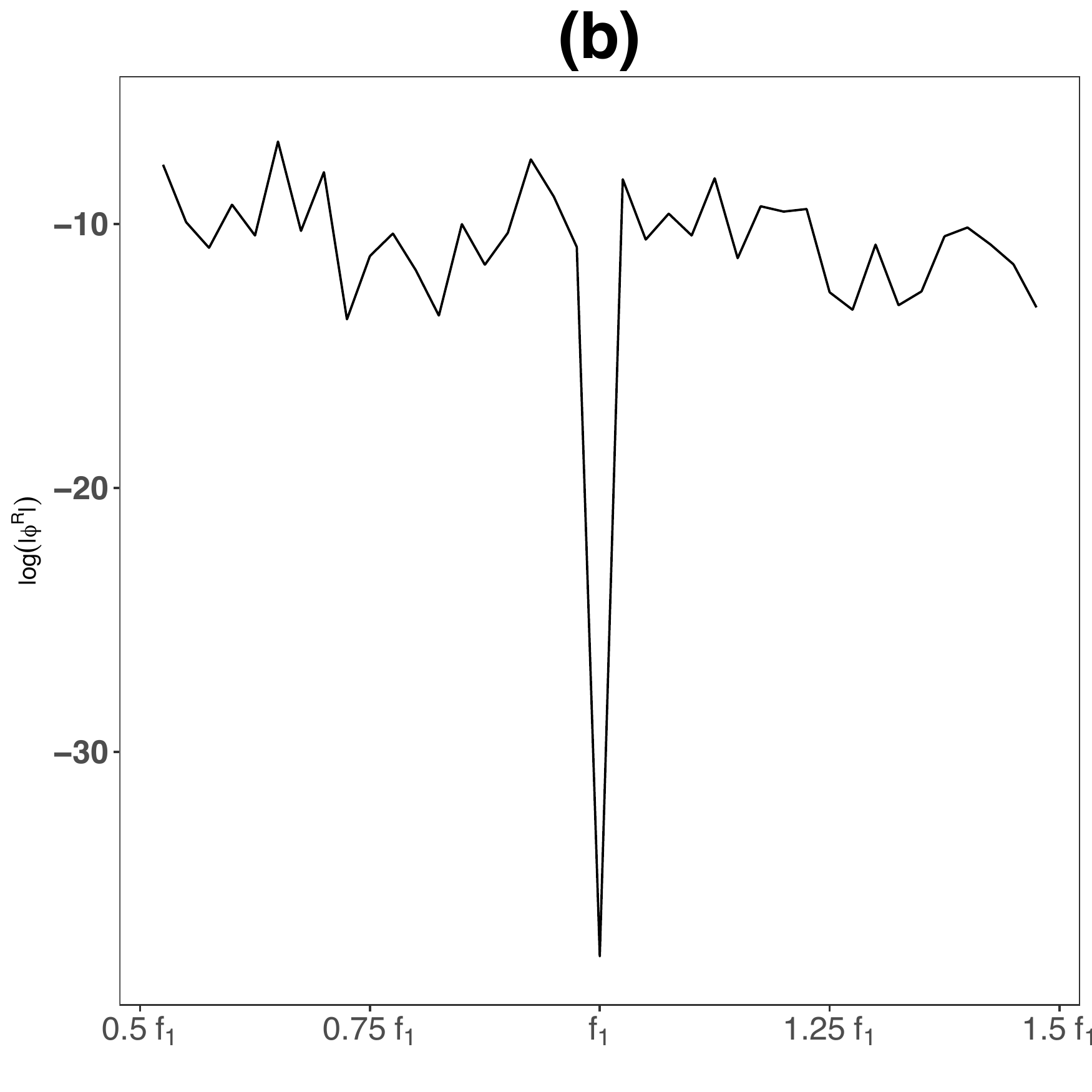} 
\end{minipage} 
\begin{minipage}{0.32\linewidth} 
\includegraphics[width=\textwidth]{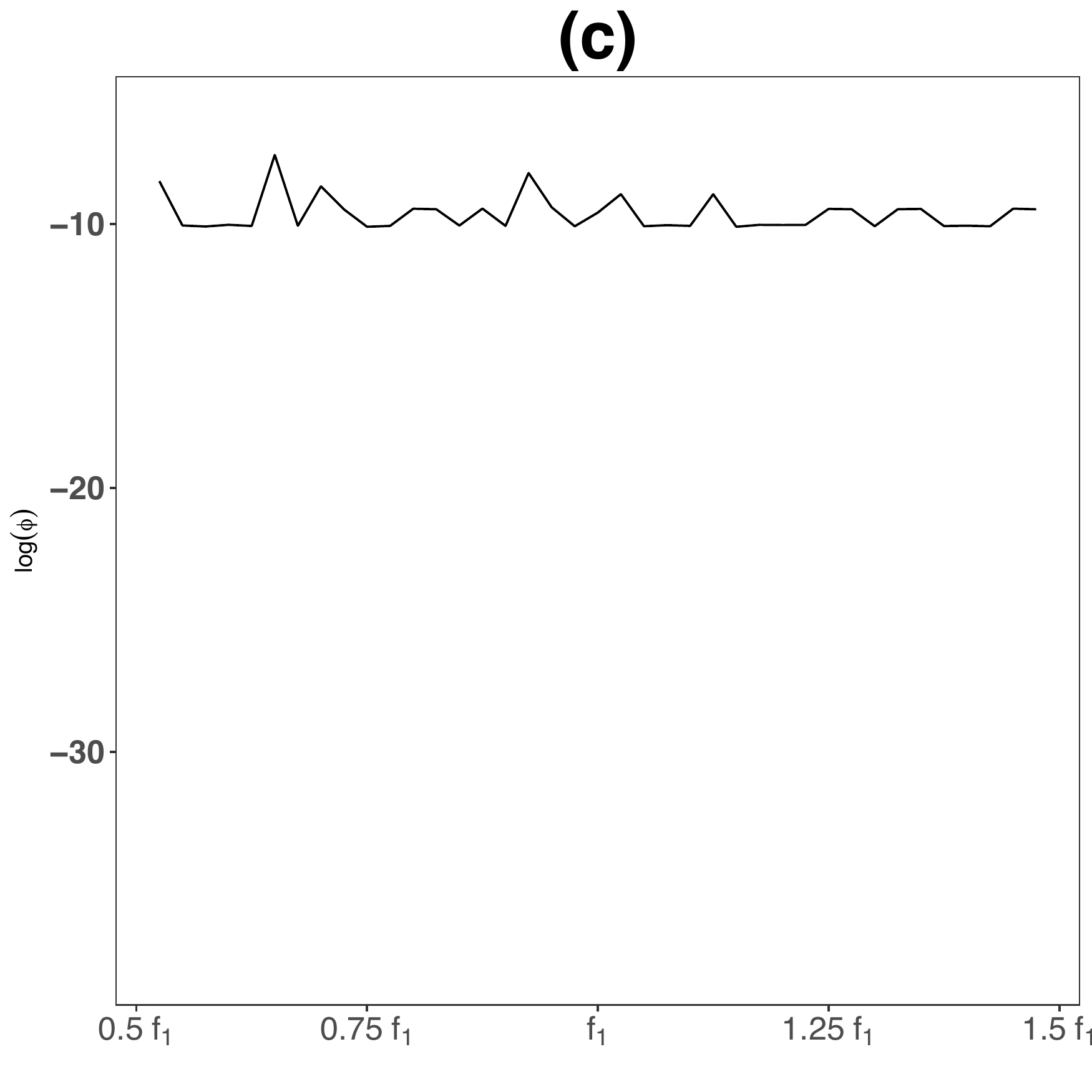} 
\end{minipage} 
\caption{a) Light curve of a RR$c$ star observed by the HIPPARCOS survey. The continuous blue line is the harmonic best fit. b) Natural logarithm  of the absolute value of the estimated parameter $\hat{\phi^R}$ by the CIAR model on the residuals of the harmonic model fitted with different frequencies;  x-axis shows percentage variation from the correct frequency, y-axis shows the natural logarithm of $\hat{\phi}^R$. c) Natural logarithm  of the estimated parameter $\hat{\phi}_{IAR}$ by the IAR model on the residuals of the harmonic model fitted with different frequencies; x-axis shows the percentual variations from the correct frequency,  y-axis  shows the natural logarithm of $\hat{\phi}_{IAR}$.\label{fig:OHex}} 
\end{figure*} 
\end{center}

Using this test,  there are some examples in which the IAR model cannot distinguish if the model is correctly fitted or not, but the CIAR model can always do this. Figure \ref{fig:OHex} shows an example of a RR$c$ star observed by the HIPPARCOS survey. Figure \eqref{fig:OHex}(a) shows the perfect fit of the harmonic model to the light curve.
 Figures \eqref{fig:OHex} (b) and (c) show that the CIAR and IAR   models, respectively, have an estimate of the parameter  close to zero on the residuals of the 39 fitted models. However, the CIAR model gives a value of $|\widehat{\phi}^R|$ significantly smaller than the remaining ones when the light curve is correctly fitted, giving a p-value close to zero (Figure \eqref{fig:OHex} (b). 

\subsection{Classification features estimated from the irregular time-series models} 
\label{ssec:Features} 

One of the most important aims in the light curve analysis from variable stars is to find features that can discriminate one class of variable star from  another. Finding a good feature is the key to building a classifier with good performance to detect stars of a given class. Generally, these features are extracted from the temporal behavior of the brightness of each variable star.\\  

In this study, we propose to use the parameter estimated by the CIAR model as a feature. For example, for a multiperiodic variable star, it can then be expected that a harmonic model fitted using only the dominant frequency will not be enough to describe its temporal dependency structure.  In other words, if we fit the CIAR and IAR models to the residuals of the harmonic model in \eqref{pArm} we should obtain values for the parameter of the ACF that are significantly different from zero. Therefore, it is natural to think that these coefficients are capable of distinguishing multiperiodic variable stars from other classes. In the OGLE and HIPPARCOS catalogs there are two classes of multiperiodic variable stars: the double-mode RR Lyraes (RRDs) and the double-mode Cepheids (DMCEPs).\\ 

We implement the IAR and the CIAR models on the residuals after fitting a harmonic model with only one period to the light curves of these classes. We found several examples of negative autocorrelation in the residuals of the harmonic model, which cannot be detected using the IAR model. One of these examples is a bi-periodic double-mode Cepheid (OGLE ID:175210) in which the IAR and the CIAR coefficients estimated on these residuals are $\hat{\phi^R}= -0.561$ and  $\hat{\phi}_{IAR}=0.011$. Consequently, we focus on the CIAR model estimates.\\

\begin{center} 
\begin{figure*}
\begin{minipage}{0.48\linewidth} 
\includegraphics[width=\textwidth]{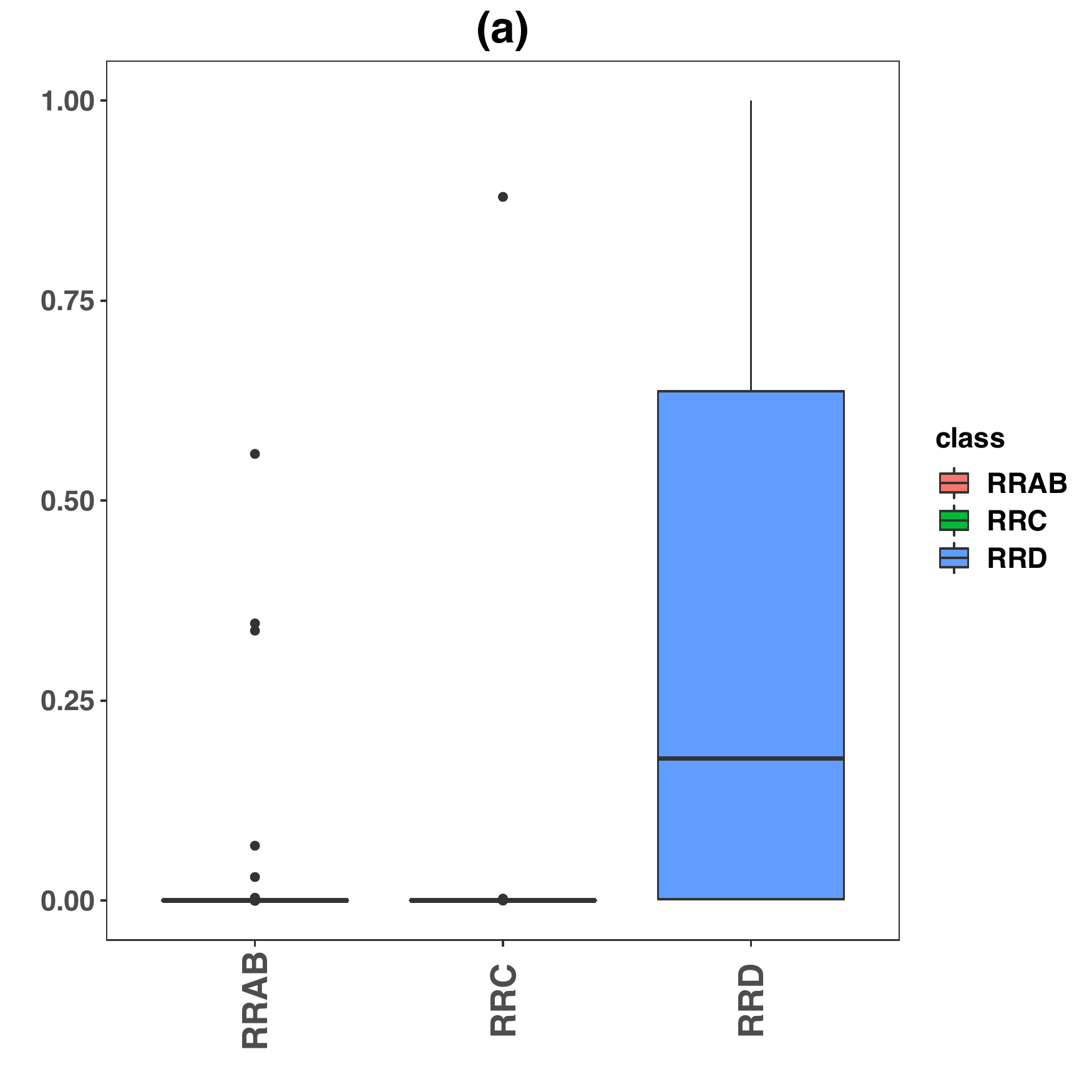} 
\end{minipage} 
\begin{minipage}{0.48\linewidth} 
\includegraphics[width=\textwidth]{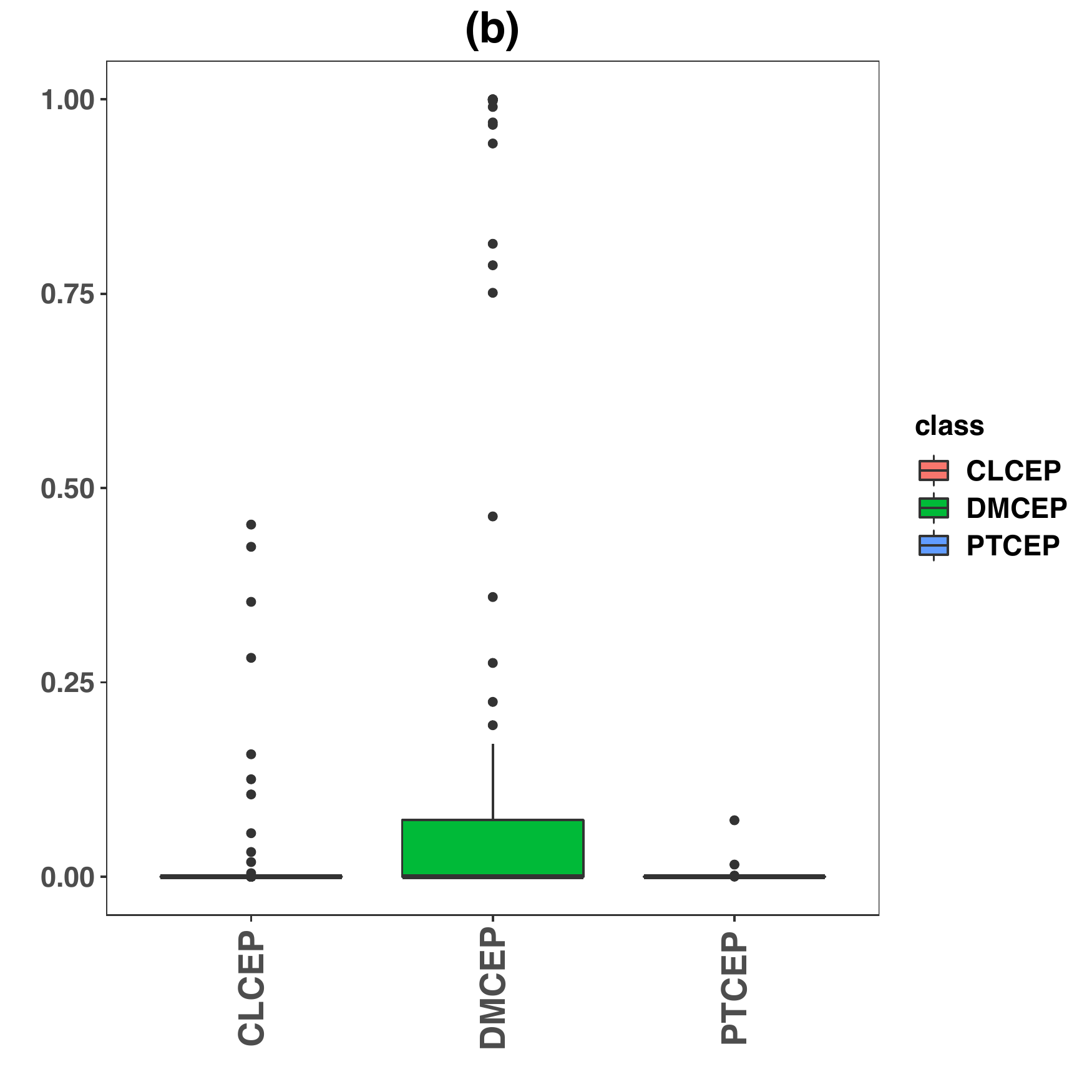} 
\end{minipage} 
\caption{a) Boxplot of the p-value estimated from the CIAR model in the RR-Lyraes variable stars separated by subclass. b) Boxplot of the p-value estimated from the CIAR model in the Cepheids variable stars separated by subclass. \label{fig:Features}} 
\end{figure*} 
\end{center}

The features that can be extracted from the CIAR model are the parameter  $\phi^R$ and the p-value associated to the test performed on  $\phi^R$. It is interesting to assess whether these features can separate the multiperiodic classes from the other RR-Lyraes and Cepheids, respectively. Figures \ref{fig:Features} (a) and (b) show the distribution of the p-values computed by the CIAR model. As can be seen from these figures  there are significant differences between the classes of RR-Lyraes and Cepheids in the distributions of the
computed  p-values. We note that the RRD and DMCEP classes take larger values in comparison to the other classes.\\

The use of the p-value as a feature reflects the multiperiodic behavior of the RRD and DMCEP classes. 

\subsection{Forecasting astronomical data} 

In this section, we illustrate the forecasting procedure implemented for the CIAR model. For this we use the light curve of the AGN MCG-6-30-15 (\cite{Lira_2015}) measured in the K-band. The time series of this object has $n=237$ observations taken over a period of approximately 4.5 years.\\

\begin{center} 
\begin{figure*}
\centering 
\begin{minipage}{0.45\linewidth} 
\includegraphics[width=\textwidth]{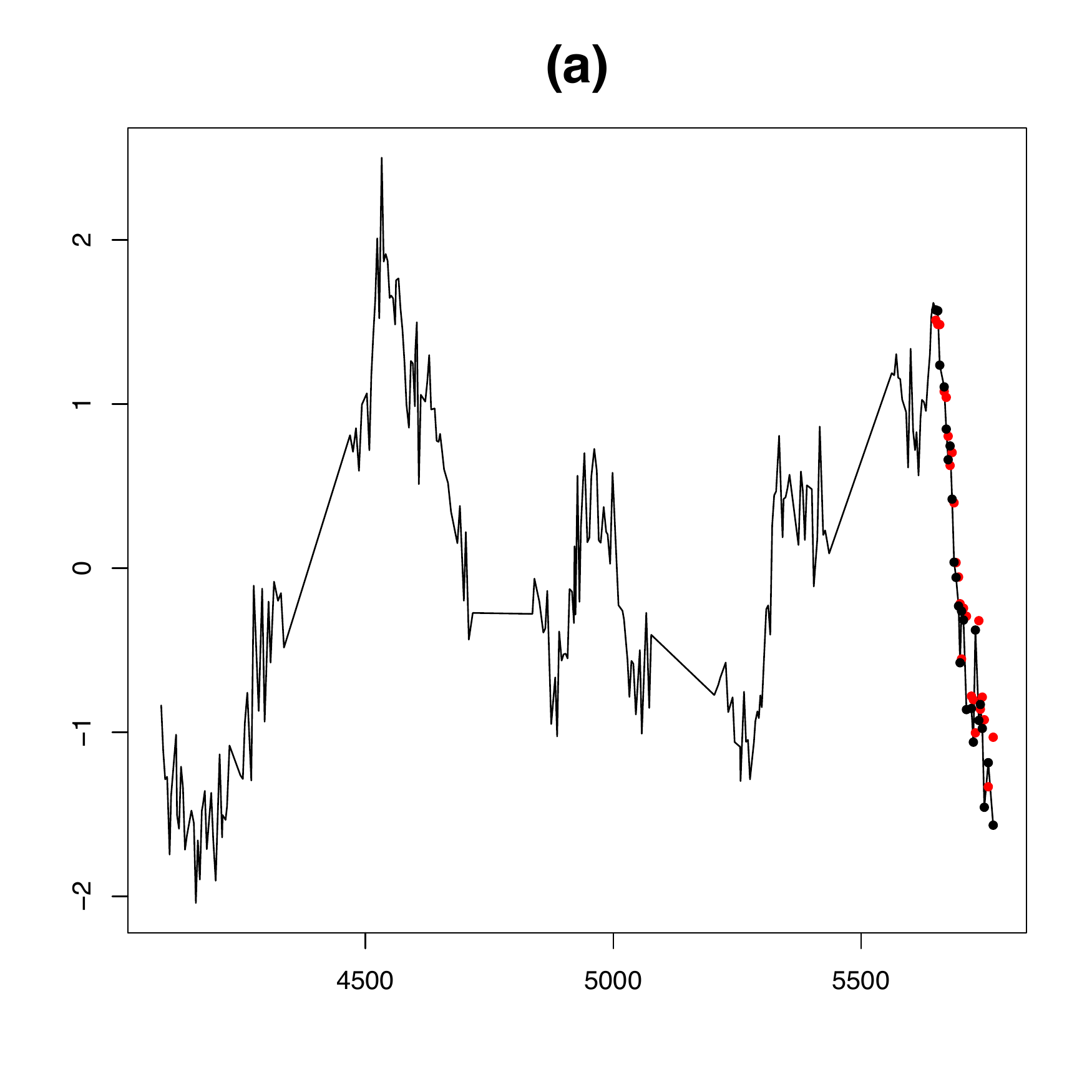} 
\end{minipage} 
\begin{minipage}{0.45\linewidth} 
\includegraphics[width=\textwidth]{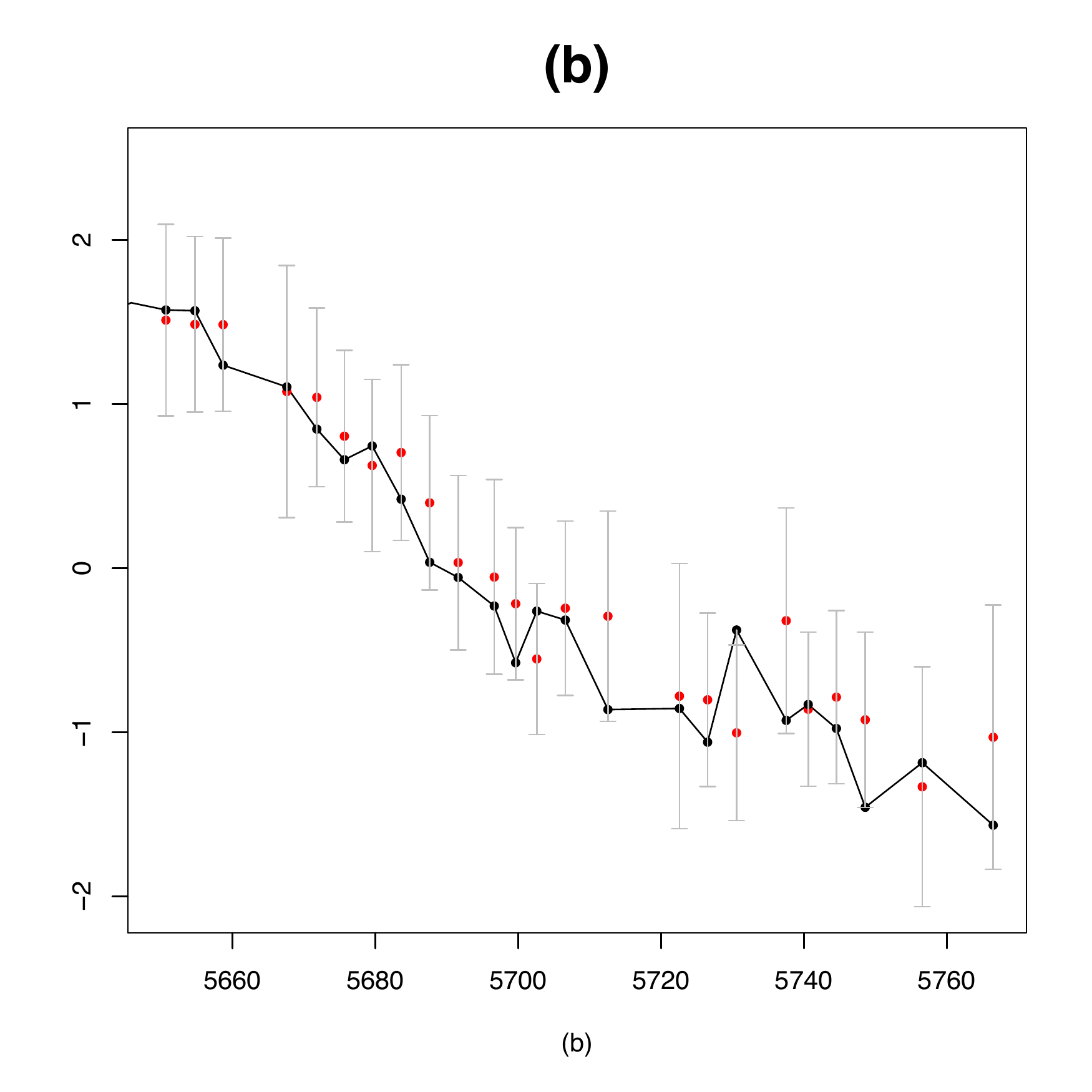} 
\end{minipage} 
\caption{a) Normalized K-band MCG-6-30-15 light curve. The red dots are the forecasted values. b) Zoom of the last 10\% of the MCG-6-30-15 light curve. The red dots are the forecasted values using the CIAR model and the gray bars are the confidence intervals at the 90\%  level.  \label{fig:AGN}} 
\end{figure*} 
\end{center}

Initially, we normalize the time series and use the first 90\% of the data to estimate the parameters of the CIAR model. We forecast the next observation at the time $t_{j+1}$ given the observational times $t_{1},\ldots,t_{j}$ corresponding to the first 90\% of the data. Later, we include the observation at the moment $t_{j+1}$ and re-estimate the model to forecast the observation at the following time $t_{j+2}$. This procedure is repeated iteratively until the remaining 10\% of the data is forecasted, obtaining the vector of one-step forecasted values $(\hat{y}_{j+1},\ldots,\hat{y}_{n})$. Figure \ref{fig:AGN} (a) shows the normalized MCG-6-30-15 light curve and the forecasted values represented with red dots.\\

The parameters estimated for this light curve were $\widehat{\phi}^R = 0.9859$ and  $\widehat{\phi}^I \approx 0$ for the first 90\% of the data and $\widehat{\phi}^R = 0.9863$ and  $\widehat{\phi}^I \approx 0$ for all the data. In addition, we also compute the confidence interval at a  90\%  level for each forecasted value, which is shown in Figure \ref{fig:AGN} (b). We note that the interval size is larger for larger time gaps.\\

%\begin{figure}
%\centering 
%\includegraphics[scale=0.45]{AGN_K.pdf} 
%\caption{One-step ahead Forecasting of AGN flux using the CIAR model \label{fig:AGN}} 
%\end{figure} 

\section{Discussion}
\label{sec:discussion}

In this work we present an extension of the irregular autoregressive (IAR) model (\cite{Eyheramendy_2018}), that we  call the { Complex Irregular Autoregressive  (CIAR) model}. As opposed to the IAR model that can only estimate positive values of the ACF, the CIAR model can estimate both positive and negative values of the ACF. We have shown that this model is weakly stationary and its state-space representation is stable under regular assumptions. We propose a maximum likelihood estimation procedure for the parameters of this model, where the solution is reached using Kalman recursions. We developed a code in R and Python to perform an estimation of the model.\\ 

There is a strong connection between the three models: CAR(1), IAR, and CIAR. Assuming a null imaginary part and positive real part of the complex autocorrelation parameter, the CIAR model becomes the IAR model. The connection between both models is verified by Monte Carlo simulations in Section \ref{ssec:Montecarlo}. Also, when Gaussian data is fitted in the models, the IAR and the CAR(1) are equivalent. Both the IAR and the CIAR models fit the irregularity of the time gaps between observations as discrete times as opposed to the CAR(1) model which considers time as a continuous variable. At this stage, only the IAR model can fit data that is not Gaussian; the CIAR and the CAR models can only fit Gaussian data. \\

We  illustrate the contribution of the CIAR model in two applications. First, the CIAR process can fit irregular time series with negative values of the  ACF, the shape of which is an exponential decay. Second, the CIAR process can identify series with negative values of the ACF such as a high frequency harmonic model or an antipersistent process. In both applications, we show with simulated data that the CIAR model performs better than other popular  time-series models.\\ 

\begin{center} 
\begin{figure*}
\begin{minipage}{0.48\linewidth} 
\includegraphics[width=\textwidth]{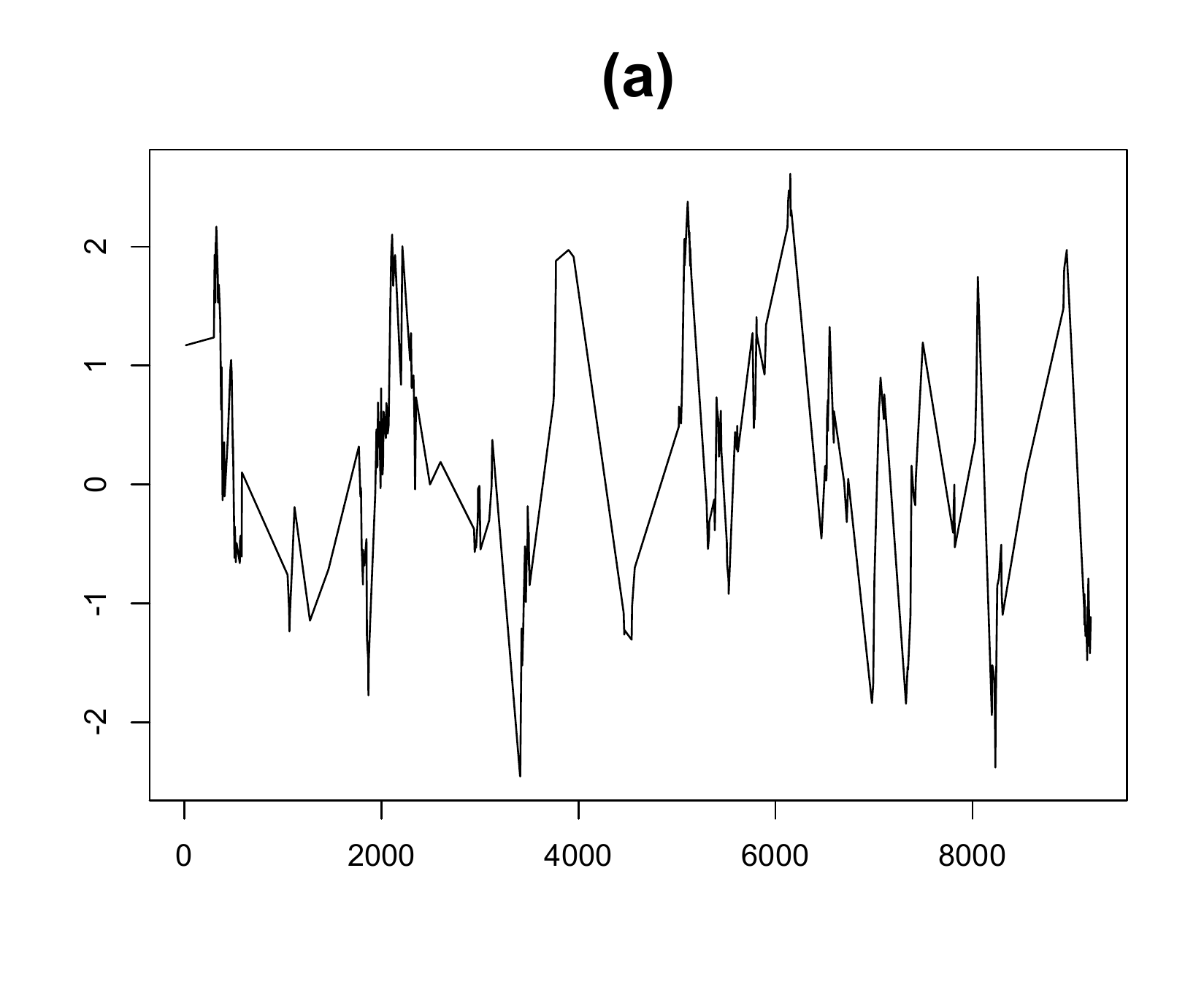} 
\end{minipage} 
\begin{minipage}{0.48\linewidth} 
\includegraphics[width=\textwidth]{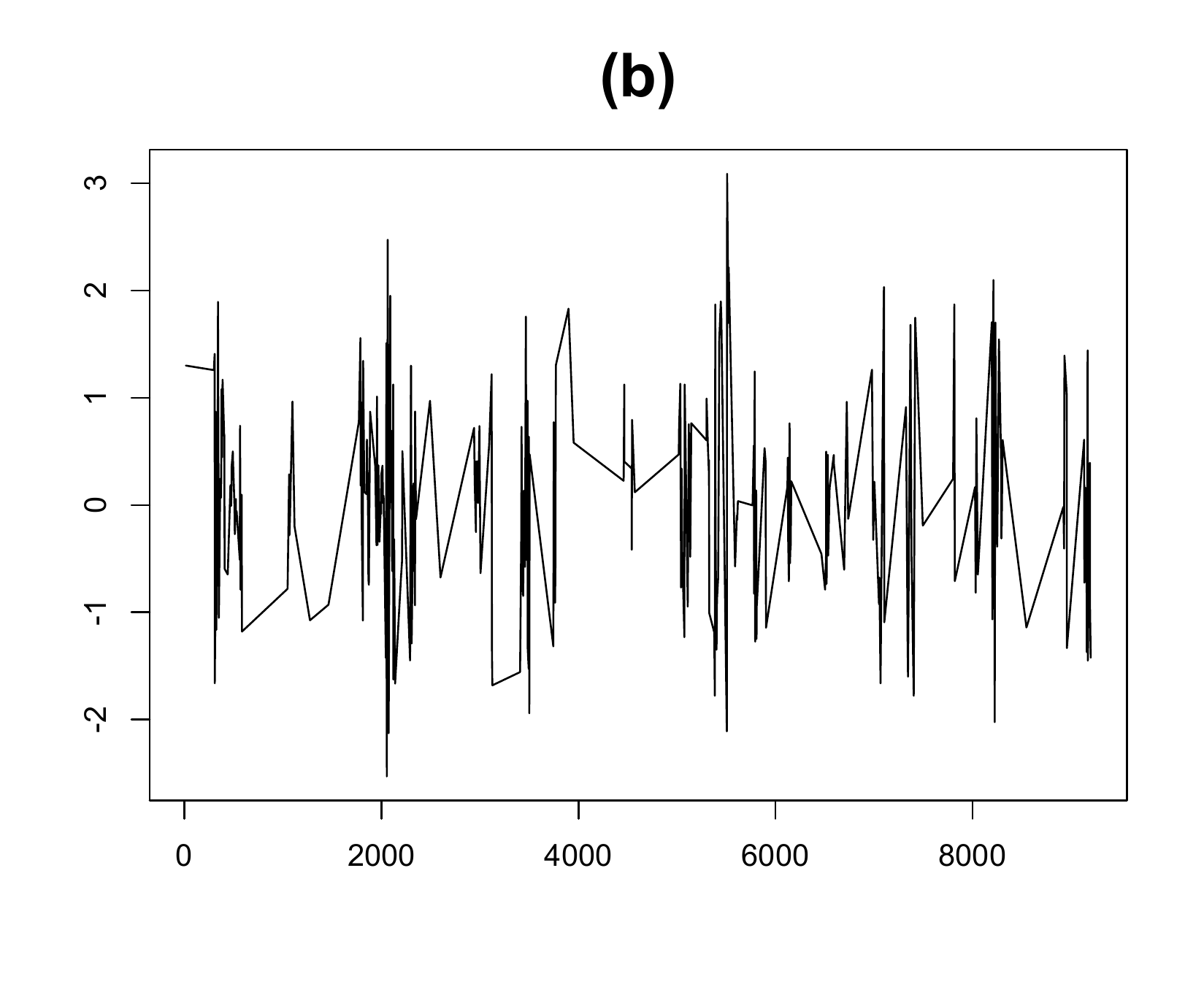} 
\end{minipage} 
\caption{a) CIAR process with positive autocorrelation generated using parameters $\phi^R = 0.99$, $\phi^I = 0$, $c=0$ and length $n=300$ . b) CIAR process with negative autocorrelation generated using parameters $\phi^R = -0.99$, $\phi^I = 0$, $c=0$ and length $n=300$. \label{fig:Negative}} 
\end{figure*} 
\end{center}

A negative autocorrelated time series is characterized by  more fluctuations  (\cite{Box_2015}). Take for example $\Delta_t=y_t-y_{t-1}$ and the variance $\mbox{Var}(\Delta_t)=2\sigma^2_y(1-\phi)$. This variance is larger for $\phi<0$ than for positive values of $\phi$. Figure \ref{fig:Negative} illustrates the behavior mentioned above from a CIAR process simulated with $\phi^R=0.99$ and $\phi^R=-0.99$. We note that there are more oscillations in the CIAR process generated with negative autocorrelation. 

The application of the proposed model is performed on astronomical data. Particularly, we use the light curves of variable stars observed by the OGLE and HIPPARCOS surveys. As many variable stars show periodical behavior in terms of their brightness, it is common to fit the light curves by a harmonic model. We have discussed that the irregular time series models are useful to determine whether or not the residuals of the harmonic fit are still autocorrelated. When a time dependency structure remains on the residuals, we can conclude either that the light curve was not correctly fitted or that it corresponds to a multiperiodic variable star. Furthermore, we noticed that the estimated coefficients are highly related to the frequency of the light curves. In order to interpret the coefficient correctly, we have developed a statistical test.\\ 

After fitting both models to the light curves, we can see a strong relationship between the IAR and CIAR models when the coefficient $\phi^R$ of the CIAR model is positive. However, we have also found several cases of negative values of the ACF of  light curves that the IAR model is not capable of identifying. In addition, we show examples where the inability to estimate negative values of $\phi$ from the IAR model prevents us from finding multiperiodic variable stars or correctly detecting whether the harmonic model is  incorrectly specified.\\ 

It is expected that negative values of the  ACF will be less frequent than  positive values.   This hypothesis is reflected in the astronomical datasets analyzed. However, there are several examples of negative values obtained from  time series of  light curves from the OGLE and HIPPARCOS surveys. This result shows how important it is to have a model that can identify the negative as well as the positive time dependencies in an irregular time series. Among the time-series models that assume irregular sampling, only the CARFIMA process can fit a negative autocorrelated time series if this has an antipersistent behavior. The contribution of the model proposed here  is that it can model series with negative exponential decay in the ACF. In other words, both processes can be negatively autocorrelated, but still have different correlation structures.\\ 

In addition, we have shown that the p-values obtained from the test proposed in this work for the CIAR model are useful for characterizing the multiperiodic classes of RR-Lyraes (RRDs) and Cepheids (DMCEPs). This result indicates that the p-value can be an important feature for a machine-learned classifier implemented on these classes. In a future study, a classifier for multiperiodic variable stars will be implemented using this feature.\\

The main aim of this work is to continue developing models for irregularly observed time series where time is considered discrete as opposed to continuous. This will allow us to extend the popular ARMA models for regular observations to the irregular case. We have shown that the CIAR model extends the IAR model by allowing negative as well as positive autocorrelations to be captured. We will continue working on expanding the scope of irregularly observed time series. \\ 

\begin{acknowledgements} Support for this research was provided by grant IC120009, awarded to The Millennium Institute of Astrophysics, MAS, and from Fondecyt grant 1160861. F.E. acknowledges support from CONICYT-PCHA (Doctorado Nacional 2014- 21140566).
\end{acknowledgements}

\bibliographystyle{aa}
\bibliography{CIARbib}

\appendix

\section{Proof of Lemma 1}
\label{sec:lem1}

Consider the CIAR process $x_{t_j}$ described by Eq. \eqref{CIAReq}. We note that $\mathbb{E}(y_{t_j}) = 0$ and $\mathbb{E}(z_{t_j}) = 0$. Therefore, $x_{t_j} = y_{t_j}+ i z_{t_j}$  is such that $\mathbb{E}(x_{t_j}) = 0$. According to the definition of the model, we have\\

\begin{eqnarray*} 
\overline{x}_{t_j}x_{t_j} &=&  \left(\overline{\phi^{\delta_{j}}} \, \overline{x}_{t_{j-1}} + \overline{\sigma}_{t_j} \, \overline{\varepsilon}_{t_j}\right)\left(\phi^{\delta_{j}} \,x_{t_{j-1}} + \sigma_{t_j} \, \varepsilon_{t_j}\right)\\ 
&=& \left|\phi^{\delta_{j}} \right|^2\left|x_{t_{j-1}} \right|^2 + \ldots + \left|\sigma_{t_j} \right|^2  \left|\varepsilon_{t_j} \right|^2. 
\end{eqnarray*} 

Applying expectation $ \mathbb{E}$ and using the properties of the model \eqref{CIAReq} 

\begin{eqnarray*} 
\gamma_0 &=&  \left|\phi^{\delta_{j}} \right|^2  \gamma_0   +  \left|\sigma_{t_j} \right|^2 (1+c),\\ 
\end{eqnarray*}

\noindent 
from which we obtain \\
\begin{eqnarray*} 
 \gamma_0 &=&  \frac{(1+c)\left|\sigma_{t_j} \right|^2}{1-\left|\phi^{\delta_{j}} \right|^2}\\ 
&=& \frac{(1+c) \sigma^2 \left( 1-\left|\phi^{\delta_{j}}\right|^2\right)}{1-\left|\phi^{\delta_{j}} \right|^2} = (1+c) \sigma^2.\\ 
\end{eqnarray*} 

The autocovariance of the process is defined by $\gamma_k = \mathbb{E}(\overline{x}_{t_{j+k}}x_{t_j})$, such that
\begin{eqnarray*} 
\mathbb{E}(\overline{x}_{t_{j+k}}x_{t_j}) &=&  \mathbb{E}\left(\left(\overline{\phi^{\delta_{j+k}}} \, \overline{x_{t_{j+k-1}}} + \overline{\sigma_{t_{j+k}}} \, \overline{\varepsilon_{t_{j+k}}}\right)x_{t_j} \right)\\ 
&=&  \overline{\phi^{\delta_{j+k}}}\mathbb{E}\left(\overline{x_{t_{j+k-1}}} x_{t_j} \right)\\ 
&=&  \overline{\phi^{\delta_{j+k}}}\mathbb{E}\left(\left(\overline{\phi^{\delta_{j+k-1}}} \, \overline{x_{t_{j+k-2}}} + \overline{\sigma_{t_{j+k-1}}} \, \overline{\varepsilon_{t_{j+k-1}}}\right) x_{t_j} \right)\\ 
&=&  \overline{\phi^{t_{j+k}-t_{j+k-2}}}\mathbb{E}\left(\overline{x_{t_{j+k-2}}}x_{t_j} \right).\\ 
\end{eqnarray*}

\noindent
It can be shown that by recursion the autocovariance function is given by 

$$\gamma_k =  \overline{\phi^{t_{j+k}-t_j}}\sigma^2 (1+c) = \overline{\phi^{\Delta_k}}\sigma^2 (1+c),\\ $$ 

\noindent
and therefore, the autocorrelation can be expressed as $\rho_k = \overline{\phi^{\Delta_k}}$ $\Box.$\\ 

\section{Proof of Lemma 2}
\label{sec:lem2}

The CIAR model is defined as $y_{t_j}+ i z_{t_j}= (\phi^R + i \phi^I)^{t_j-t_{j-1}} \, (y_{t_{j-1}} + i z_{t_{j-1}}) + \sigma_{t_j}(\varepsilon_{t_j}^R + i \varepsilon_{t_j}^I)$. Let us focus on the term $ (\phi^R + i \phi^I)^{t_j-t_{j-1}}$,

\begin{eqnarray*}
 (\phi^R + i \phi^I)^{t_j-t_{j-1}} &=&  (\phi^R + i \phi^I)^{\delta_{j}} = |\phi|^{\delta_{j}} \left(\frac{\phi^R + i \phi^I}{|\phi|} \right)^{\delta_{j}}.
\end{eqnarray*}

\noindent
 Using the polar representation for complex numbers we obtain

\begin{eqnarray*}
\left(\frac{\phi^R + i \phi^I}{|\phi|} \right)^{\delta_{j}} &=& (\cos(\psi) + i \sin(\psi))^{\delta_{j}}\\
\left(\phi^R + i \phi^I\right)^{\delta_{j}} &=& |\phi|^{\delta_{j}} (\cos(\psi) + i \sin(\psi))^{\delta_{j}}.
\end{eqnarray*}

\noindent
 Using the De Moivre formula, we have
\begin{eqnarray*}
\left(\phi^R + i \phi^I\right)^{\delta_{j}} &=& |\phi|^{\delta_{j}} (\cos({\delta_{j}} \psi) + i \sin({\delta_{j}} \psi))\\
&=& |\phi|^{\delta_{j}} \cos({\delta_{j}} \psi) + i  |\phi|^{\delta_{j}} \sin({\delta_{j}} \psi) \\
&=& \alpha_{t_j}^{R} + i \alpha_{t_j}^{I}.\\
\end{eqnarray*}

\noindent
 Finally, the Complex IAR model can be represented by the expression

\begin{eqnarray*}
y_{t_j}+ i z_{t_j}= (\alpha_{t_j}^{R} + i \alpha_{t_j}^{I}) \, (y_{t_{j-1}} + i z_{t_{j-1}}) + \sigma_{t_j}(\varepsilon_{t_j}^R + i \varepsilon_{t_j}^I) \Box.\\
\end{eqnarray*}

\section{Proof of Lemma 3}
\label{sec:lem3} 

At a fixed time $t_j$, the eigenvalues of the transition matrix of the CIAR process $F_{t_j}  =  \left(\begin{array}{cc} \alpha_{t_j}^{R}  & -\alpha_{t_j}^{I} \\ \alpha_{t_j}^{I} & \alpha_{t_j}^{R} \end{array} \right)$ satisfy the following equation.

\begin{eqnarray*}
|(F_{t_j} -\lambda I)| &=& \left|  \left(\begin{array}{cc} \alpha_{t_j}^{R}  & -\alpha_{t_j}^{I} \\ \alpha_{t_j}^{I} & \alpha_{t_j}^{R} \end{array} \right) - \left(\begin{array}{cc} \lambda  & 0 \\ 0 & \lambda \end{array} \right)\right| \\
&=& \left| \left(\begin{array}{cc} \alpha_{t_j}^{R} - \lambda  & -\alpha_{t_j}^{I} \\ \alpha_{t_j}^{I} & \alpha_{t_j}^{R} - \lambda \end{array} \right)\right|\\
 &=&  (\alpha_{t_j}^{R} - \lambda)^2 + \alpha_{t_j}^{I2} \\
 &=&  \lambda^2 - 2 \alpha_{t_j}^{R} \lambda + (\alpha_{t_j}^{R2}+\alpha_{t_j}^{I2}) \\
 &=& 0,\\
\end{eqnarray*}

$\Rightarrow \lambda = \frac{2\alpha_{t_j}^{R} \pm \sqrt{4\alpha_{t_j}^{R2} -  4(\alpha_{t_j}^{R2}+\alpha_{t_j}^{I2})}}{2} = \alpha_{t_j}^R \pm i \alpha_{t_j}^{I}$.\\

Since $ |\alpha_{t_j}^R + i \alpha_{t_j}^{I}| = |\alpha_{t_j}^R - i \alpha_{t_j}^{I}|  = |\alpha_{t_j}|$, then the process is stable if $\sup |\alpha_{t_j}| < 1$. Therefore, under this assumption the CIAR process has the unique stationary solution (\cite{Brockwell_2002}) given by
\begin{eqnarray*}
X_{t_j} &=& F_{t_j} X_{t_{j-1}}  + V_{t_j} \\
&=& F_{t_j} (F_{t_{j-1}} X_{t_{j-2}}  + V_{t_{j-1}})  + V_{t_j} \\
&=& F_{t_j} F_{t_{j-1}} X_{t_{j-2}}  + F_{t_j}V_{t_{j-1}}  + V_{t_j}\\
&=& F_{t_j} F_{t_{j-1}}  (F_{t_{j-2}} X_{t_{j-3}}  + V_{t_{j-2}})  + F_{t_j}V_{t_{j-1}}  + V_{t_j} \\
&=& F_{t_j} F_{t_{j-1}}  F_{t_{j-2}} X_{t_{j-3}}  +  F_{t_j} F_{t_{j-1}} V_{t_{j-2}} \\
&+& F_{t_j}V_{t_{j-1}}  + V_{t_j}.
\end{eqnarray*}

Therefore, the general form can be written as,

$$X_{t_j} = X_{t_{j-n}}\mathop{\prod}_{k=0}^{n-1} F_{t_{j-k}} + V_{t_j} + \mathop{\sum}_{k=1}^{n-1} V_{t_{j-k}} \mathop{\prod}_{i=0}^{k-1} F_{t_{j-i}}.$$

Since $\left|\mathop{\prod}_{k=0}^{n-1} F_{t_{j-k}} \right| = \mathop{\prod}_{k=0}^{n-1} \left| F_{t_{j-k}} \right|$ and $|F_{t_{j-k}}| < 1$ due to the stability of the process, then $\lim\limits_{n\to\infty} \mathop{\prod}_{k=0}^{n-1} F_{t_{j-k}} = 0$. Finally, if $n\to\infty$ then the unique stationary solution is given by,

$$X_{t_j} = V_{t_j} + \mathop{\sum}_{k=1}^{\infty} V_{t_{j-k}} \mathop{\prod}_{i=0}^{k-1} F_{t_{j-i}} \Box.$$

\end{document}